\ifx\mnmacrosloaded\undefined 
%
%
%
%

\catcode `\@=11 

\def\@version{1.6}
\def\@verdate{18th September 1995}

%
%


\newif\ifprod@font

\ifx\@typeface\undefined
  \def\@typeface{Comp. Modern}\prod@fontfalse
\else
  \prod@fonttrue 
\fi

\def\newfam{\alloc@8\fam\chardef\sixt@@n} 

\ifprod@font
\font\fiverm=mtr10 at 5pt
\font\fivebf=mtbx10 at 5pt
\font\fiveit=mtti10 at 5pt
\font\fivesl=mtsl10 at 5pt
\font\fivett=cmtt8 at 5pt     \hyphenchar\fivett=-1
\font\fivecsc=mtcsc10 at 5pt
\font\fivesf=mtss10 at 5pt
\font\fivei=mtmi10 at 5pt      \skewchar\fivei='177
\font\fivesy=mtsy10 at 5pt     \skewchar\fivesy='60

\font\sixrm=mtr10 at 6pt
\font\sixbf=mtbx10 at 6pt
\font\sixit=mtti10 at 6pt
\font\sixsl=mtsl10 at 6pt
\font\sixtt=cmtt8 at 6pt      \hyphenchar\sixtt=-1
\font\sixcsc=mtcsc10 at 6pt
\font\sixsf=mtss10 at 6pt
\font\sixi=mtmi10 at 6pt       \skewchar\sixi='177
\font\sixsy=mtsy10 at 6pt      \skewchar\sixsy='60

\font\sevenrm=mtr10 at 7pt
\font\sevenbf=mtbx10 at 7pt
\font\sevenit=mtti10 at 7pt
\font\sevensl=mtsl10 at 7pt
\font\seventt=cmtt8 at 7pt     \hyphenchar\seventt=-1
\font\sevencsc=mtcsc10 at 7pt
\font\sevensf=mtss10 at 7pt
\font\seveni=mtmi10 at 7pt      \skewchar\seveni='177
\font\sevensy=mtsy10 at 7pt     \skewchar\sevensy='60

\font\eightrm=mtr10 at 8pt
\font\eightbf=mtbx10 at 8pt
\font\eightit=mtti10 at 8pt
\font\eighti=mtmi10 at 8pt      \skewchar\eighti='177
\font\eightsy=mtsy10 at 8pt     \skewchar\eightsy='60
\font\eightsl=mtsl10 at 8pt
\font\eighttt=cmtt8             \hyphenchar\eighttt=-1
\font\eightcsc=mtcsc10 at 8pt
\font\eightsf=mtss10 at 8pt

\font\ninerm=mtr10 at 9pt
\font\ninebf=mtbx10 at 9pt
\font\nineit=mtti10 at 9pt
\font\ninei=mtmi10 at 9pt      \skewchar\ninei='177
\font\ninesy=mtsy10 at 9pt     \skewchar\ninesy='60
\font\ninesl=mtsl10 at 9pt
\font\ninett=cmtt9             \hyphenchar\ninett=-1
\font\ninecsc=mtcsc10 at 9pt
\font\ninesf=mtss10 at 9pt

\font\tenrm=mtr10
\font\tenbf=mtbx10
\font\tenit=mtti10
\font\teni=mtmi10		\skewchar\teni='177
\font\tensy=mtsy10		\skewchar\tensy='60
\font\tenex=cmex10
\font\tensl=mtsl10
\font\tentt=cmtt10		\hyphenchar\tentt=-1
\font\tencsc=mtcsc10
\font\tensf=mtss10

\font\elevenrm=mtr10 at 11pt
\font\elevenbf=mtbx10 at 11pt
\font\elevenit=mtti10 at 11pt
\font\eleveni=mtmi10 at 11pt      \skewchar\eleveni='177
\font\elevensy=mtsy10 at 11pt     \skewchar\elevensy='60
\font\elevensl=mtsl10 at 11pt
\font\eleventt=cmtt10 at 11pt     \hyphenchar\eleventt=-1
\font\elevencsc=mtcsc10 at 11pt
\font\elevensf=mtss10 at 11pt

\font\twelverm=mtr10 at 12pt
\font\twelvebf=mtbx10 at 12pt
\font\twelveit=mtti10 at 12pt
\font\twelvesl=mtsl10 at 12pt
\font\twelvett=cmtt12             \hyphenchar\twelvett=-1
\font\twelvecsc=mtcsc10 at 12pt
\font\twelvesf=mtss10 at 12pt
\font\twelvei=mtmi10 at 12pt      \skewchar\twelvei='177
\font\twelvesy=mtsy10 at 12pt     \skewchar\twelvesy='60

\font\fourteenrm=mtr10 at 14pt
\font\fourteenbf=mtbx10 at 14pt
\font\fourteenit=mtti10 at 14pt
\font\fourteeni=mtmi10 at 14pt      \skewchar\fourteeni='177
\font\fourteensy=mtsy10 at 14pt     \skewchar\fourteensy='60
\font\fourteensl=mtsl10 at 14pt
\font\fourteentt=cmtt12 at 14pt     \hyphenchar\fourteentt=-1
\font\fourteencsc=mtcsc10 at 14pt
\font\fourteensf=mtss10 at 14pt

\font\seventeenrm=mtr10 at 17pt
\font\seventeenbf=mtbx10 at 17pt
\font\seventeenit=mtti10 at 17pt
\font\seventeeni=mtmi10 at 17pt      \skewchar\seventeeni='177
\font\seventeensy=mtsy10 at 17pt     \skewchar\seventeensy='60
\font\seventeensl=mtsl10 at 17pt
\font\seventeentt=cmtt12 at 17pt     \hyphenchar\seventeentt=-1
\font\seventeencsc=mtcsc10 at 17pt
\font\seventeensf=mtss10 at 17pt
\else
\font\fiverm=cmr5
\font\fivei=cmmi5             \skewchar\fivei='177
\font\fivesy=cmsy5            \skewchar\fivesy='60
\font\fivebf=cmbx5

\font\sixrm=cmr6
\font\sixi=cmmi6             \skewchar\sixi='177
\font\sixsy=cmsy6            \skewchar\sixsy='60
\font\sixbf=cmbx6

\font\sevenrm=cmr7
\font\sevenit=cmti7
\font\seveni=cmmi7             \skewchar\seveni='177
\font\sevensy=cmsy7            \skewchar\sevensy='60
\font\sevenbf=cmbx7

\font\eightrm=cmr8
\font\eightbf=cmbx8
\font\eightit=cmti8
\font\eighti=cmmi8			\skewchar\eighti='177
\font\eightsy=cmsy8			\skewchar\eightsy='60
\font\eightsl=cmsl8
\font\eighttt=cmtt8			\hyphenchar\eighttt=-1
\font\eightcsc=cmcsc10 at 8pt
\font\eightsf=cmss8

\font\ninerm=cmr9
\font\ninebf=cmbx9
\font\nineit=cmti9
\font\ninei=cmmi9			\skewchar\ninei='177
\font\ninesy=cmsy9			\skewchar\ninesy='60
\font\ninesl=cmsl9
\font\ninett=cmtt9			\hyphenchar\ninett=-1
\font\ninecsc=cmcsc10 at 9pt
\font\ninesf=cmss9

\font\tenrm=cmr10
\font\tenbf=cmbx10
\font\tenit=cmti10
\font\teni=cmmi10		\skewchar\teni='177
\font\tensy=cmsy10		\skewchar\tensy='60
\font\tenex=cmex10
\font\tensl=cmsl10
\font\tentt=cmtt10		\hyphenchar\tentt=-1
\font\tencsc=cmcsc10
\font\tensf=cmss10

\font\elevenrm=cmr10 scaled \magstephalf
\font\elevenbf=cmbx10 scaled \magstephalf
\font\elevenit=cmti10 scaled \magstephalf
\font\eleveni=cmmi10 scaled \magstephalf	\skewchar\eleveni='177
\font\elevensy=cmsy10 scaled \magstephalf	\skewchar\elevensy='60
\font\elevensl=cmsl10 scaled \magstephalf
\font\eleventt=cmtt10 scaled \magstephalf	\hyphenchar\eleventt=-1
\font\elevencsc=cmcsc10 scaled \magstephalf
\font\elevensf=cmss10 scaled \magstephalf

\font\twelverm=cmr10 scaled \magstep1
\font\twelvebf=cmbx10 scaled \magstep1
\font\twelvei=cmmi10 scaled \magstep1      \skewchar\twelvei='177
\font\twelvesy=cmsy10 scaled \magstep1     \skewchar\twelvesy='60

\font\fourteenrm=cmr10 scaled \magstep2
\font\fourteenbf=cmbx10 scaled \magstep2
\font\fourteenit=cmti10 scaled \magstep2
\font\fourteeni=cmmi10 scaled \magstep2		\skewchar\fourteeni='177
\font\fourteensy=cmsy10 scaled \magstep2	\skewchar\fourteensy='60
\font\fourteensl=cmsl10 scaled \magstep2
\font\fourteentt=cmtt10 scaled \magstep2	\hyphenchar\fourteentt=-1
\font\fourteencsc=cmcsc10 scaled \magstep2
\font\fourteensf=cmss10 scaled \magstep2

\font\seventeenrm=cmr10 scaled \magstep3
\font\seventeenbf=cmbx10 scaled \magstep3
\font\seventeenit=cmti10 scaled \magstep3
\font\seventeeni=cmmi10 scaled \magstep3	\skewchar\seventeeni='177
\font\seventeensy=cmsy10 scaled \magstep3	\skewchar\seventeensy='60
\font\seventeensl=cmsl10 scaled \magstep3
\font\seventeentt=cmtt10 scaled \magstep3	\hyphenchar\seventeentt=-1
\font\seventeencsc=cmcsc10 scaled \magstep3
\font\seventeensf=cmss10 scaled \magstep3
\fi

\def\hexnumber#1{\ifcase#1 0\or1\or2\or3\or4\or5\or6\or7\or8\or9\or
  A\or B\or C\or D\or E\or F\fi}

\def\makestrut{%
  \setbox\strutbox=\hbox{%
    \vrule height.7\baselineskip depth.3\baselineskip width \z@}%
}

\def\baselinestretch{1}
\newskip\tmp@bls

\def\b@ls#1{
  \tmp@bls=#1\relax
  \baselineskip=#1\relax\makestrut
  \normalbaselineskip=\baselinestretch\tmp@bls
  \normalbaselines
}

\def\nostb@ls#1{
  \normalbaselineskip=#1\relax
  \normalbaselines
  \makestrut
}

%

\newfam\scfam  
\newfam\sffam  

\def\mit{\fam\@ne}
\def\cal{\fam\tw@}
\def\em{\ifdim\fontdimen1\font>\z@ \rm\else\it\fi}

\textfont3=\tenex
\scriptfont3=\tenex
\scriptscriptfont3=\tenex

\setbox0=\hbox{\tenex B} \p@renwd=\wd0 

\def\eightpoint{
  \def\rm{\fam0\eightrm}%
  \textfont0=\eightrm \scriptfont0=\sixrm \scriptscriptfont0=\fiverm%
  \textfont1=\eighti  \scriptfont1=\sixi  \scriptscriptfont1=\fivei%
  \textfont2=\eightsy \scriptfont2=\sixsy \scriptscriptfont2=\fivesy%
  \textfont\itfam=\eightit\def\it{\fam\itfam\eightit}%
  \ifprod@font
    \scriptfont\itfam=\sixit
      \scriptscriptfont\itfam=\fiveit
  \else
    \scriptfont\itfam=\eightit
      \scriptscriptfont\itfam=\eightit
  \fi
  \textfont\bffam=\eightbf%
    \scriptfont\bffam=\sixbf%
      \scriptscriptfont\bffam=\fivebf%
  \def\bf{\fam\bffam\eightbf}%
  \textfont\slfam=\eightsl\def\sl{\fam\slfam\eightsl}%
  \ifprod@font
    \scriptfont\slfam=\sixsl
      \scriptscriptfont\slfam=\fivesl
  \else
    \scriptfont\slfam=\eightsl
      \scriptscriptfont\slfam=\eightsl
  \fi
  \textfont\ttfam=\eighttt\def\tt{\fam\ttfam\eighttt}%
  \ifprod@font
    \scriptfont\ttfam=\sixtt
      \scriptscriptfont\ttfam=\fivett
  \else
    \scriptfont\ttfam=\eighttt
      \scriptscriptfont\ttfam=\eighttt
  \fi
  \textfont\scfam=\eightcsc\def\sc{\fam\scfam\eightcsc}%
  \ifprod@font
    \scriptfont\scfam=\sixcsc
      \scriptscriptfont\scfam=\fivecsc
  \else
    \scriptfont\scfam=\eightcsc
      \scriptscriptfont\scfam=\eightcsc
  \fi
  \textfont\sffam=\eightsf\def\sf{\fam\sffam\eightsf}%
  \ifprod@font
    \scriptfont\sffam=\sixsf
      \scriptscriptfont\sffam=\fivesf
  \else
    \scriptfont\sffam=\eightsf
      \scriptscriptfont\sffam=\eightsf
  \fi
  \def\oldstyle{\fam\@ne\eighti}%
  \b@ls{10pt}\rm\@viiipt%
}
\def\@viiipt{}

\def\ninepoint{
  \def\rm{\fam0\ninerm}%
  \textfont0=\ninerm \scriptfont0=\sixrm \scriptscriptfont0=\fiverm%
  \textfont1=\ninei  \scriptfont1=\sixi  \scriptscriptfont1=\fivei%
  \textfont2=\ninesy \scriptfont2=\sixsy \scriptscriptfont2=\fivesy%
  \textfont\itfam=\nineit\def\it{\fam\itfam\nineit}%
  \ifprod@font
    \scriptfont\itfam=\sixit
      \scriptscriptfont\itfam=\fiveit
  \else
    \scriptfont\itfam=\nineit
      \scriptscriptfont\itfam=\nineit
  \fi
  \textfont\bffam=\ninebf%
    \scriptfont\bffam=\sixbf%
      \scriptscriptfont\bffam=\fivebf%
  \def\bf{\fam\bffam\ninebf}%
  \textfont\slfam=\ninesl\def\sl{\fam\slfam\ninesl}%
  \ifprod@font
    \scriptfont\slfam=\sixsl
      \scriptscriptfont\slfam=\fivesl
  \else
    \scriptfont\slfam=\ninesl
      \scriptscriptfont\slfam=\ninesl
  \fi
  \textfont\ttfam=\ninett\def\tt{\fam\ttfam\ninett}%
  \ifprod@font
    \scriptfont\ttfam=\sixtt
      \scriptscriptfont\ttfam=\fivett
  \else
    \scriptfont\ttfam=\ninett
      \scriptscriptfont\ttfam=\ninett
  \fi
  \textfont\scfam=\ninecsc\def\sc{\fam\scfam\ninecsc}%
  \ifprod@font
    \scriptfont\scfam=\sixcsc
      \scriptscriptfont\scfam=\fivecsc
  \else
    \scriptfont\scfam=\ninecsc
      \scriptscriptfont\scfam=\ninecsc
  \fi
  \textfont\sffam=\ninesf\def\sf{\fam\sffam\ninesf}%
  \ifprod@font
    \scriptfont\sffam=\sixsf
      \scriptscriptfont\sffam=\fivesf
  \else
    \scriptfont\sffam=\ninesf
      \scriptscriptfont\sffam=\ninesf
  \fi
  \def\oldstyle{\fam\@ne\ninei}%
  \b@ls{\TextLeading plus \Feathering}\rm\@ixpt%
}
\def\@ixpt{}

\def\tenpoint{
  \def\rm{\fam0\tenrm}%
  \textfont0=\tenrm \scriptfont0=\sevenrm \scriptscriptfont0=\fiverm%
  \textfont1=\teni  \scriptfont1=\seveni  \scriptscriptfont1=\fivei%
  \textfont2=\tensy \scriptfont2=\sevensy \scriptscriptfont2=\fivesy%
  \textfont\itfam=\tenit\def\it{\fam\itfam\tenit}%
  \ifprod@font
    \scriptfont\itfam=\sevenit
      \scriptscriptfont\itfam=\fiveit
  \else
    \scriptfont\itfam=\tenit
      \scriptscriptfont\itfam=\tenit
  \fi
  \textfont\bffam=\tenbf%
    \scriptfont\bffam=\sevenbf%
      \scriptscriptfont\bffam=\fivebf%
  \def\bf{\fam\bffam\tenbf}%
  \textfont\slfam=\tensl\def\sl{\fam\slfam\tensl}%
  \ifprod@font
    \scriptfont\slfam=\sevensl
      \scriptscriptfont\slfam=\fivesl
  \else
    \scriptfont\slfam=\tensl
      \scriptscriptfont\slfam=\tensl
  \fi
  \textfont\ttfam=\tentt\def\tt{\fam\ttfam\tentt}%
  \ifprod@font
    \scriptfont\ttfam=\seventt
      \scriptscriptfont\ttfam=\fivett
  \else
    \scriptfont\ttfam=\tentt
      \scriptscriptfont\ttfam=\tentt
  \fi
  \textfont\scfam=\tencsc\def\sc{\fam\scfam\tencsc}%
  \ifprod@font
    \scriptfont\scfam=\sevencsc
      \scriptscriptfont\scfam=\fivecsc
  \else
    \scriptfont\scfam=\tencsc
      \scriptscriptfont\scfam=\tencsc
  \fi
  \textfont\sffam=\tensf\def\sf{\fam\sffam\tensf}%
  \ifprod@font
    \scriptfont\sffam=\sevensf
      \scriptscriptfont\sffam=\fivesf
  \else
    \scriptfont\sffam=\tensf
      \scriptscriptfont\sffam=\tensf
  \fi
  \def\oldstyle{\fam\@ne\teni}%
  \b@ls{11pt}\rm\@xpt%
}
\def\@xpt{}

\def\elevenpoint{
  \def\rm{\fam0\elevenrm}%
  \textfont0=\elevenrm \scriptfont0=\eightrm \scriptscriptfont0=\sixrm%
  \textfont1=\eleveni  \scriptfont1=\eighti  \scriptscriptfont1=\sixi%
  \textfont2=\elevensy \scriptfont2=\eightsy \scriptscriptfont2=\sixsy%
  \textfont\itfam=\elevenit\def\it{\fam\itfam\elevenit}%
  \ifprod@font
    \scriptfont\itfam=\eightit
      \scriptscriptfont\itfam=\sixit
  \else
    \scriptfont\itfam=\elevenit
      \scriptscriptfont\itfam=\elevenit
  \fi
  \textfont\bffam=\elevenbf%
    \scriptfont\bffam=\eightbf%
      \scriptscriptfont\bffam=\sixbf%
  \def\bf{\fam\bffam\elevenbf}%
  \textfont\slfam=\elevensl\def\sl{\fam\slfam\elevensl}%
  \ifprod@font
    \scriptfont\slfam=\eightsl
      \scriptscriptfont\slfam=\sixsl
  \else
    \scriptfont\slfam=\elevensl
      \scriptscriptfont\slfam=\elevensl
  \fi
  \textfont\ttfam=\eleventt\def\tt{\fam\ttfam\eleventt}%
  \ifprod@font
    \scriptfont\ttfam=\eighttt
      \scriptscriptfont\ttfam=\sixtt
  \else
    \scriptfont\ttfam=\eleventt
      \scriptscriptfont\ttfam=\eleventt
  \fi
  \textfont\scfam=\elevencsc\def\sc{\fam\scfam\elevencsc}%
  \ifprod@font
    \scriptfont\scfam=\eightcsc
      \scriptscriptfont\scfam=\sixcsc
  \else
    \scriptfont\scfam=\elevencsc
      \scriptscriptfont\scfam=\elevencsc
  \fi
  \textfont\sffam=\elevensf\def\sf{\fam\sffam\elevensf}%
  \ifprod@font
    \scriptfont\sffam=\eightsf
      \scriptscriptfont\sffam=\sixsf
  \else
    \scriptfont\sffam=\elevensf
      \scriptscriptfont\sffam=\elevensf
  \fi
  \def\oldstyle{\fam\@ne\eleveni}%
  \b@ls{13pt}\rm\@xipt%
}
\def\@xipt{}

\def\fourteenpoint{
  \def\rm{\fam0\fourteenrm}%
  \textfont0\fourteenrm  \scriptfont0\tenrm  \scriptscriptfont0\sevenrm%
  \textfont1\fourteeni   \scriptfont1\teni   \scriptscriptfont1\seveni%
  \textfont2\fourteensy  \scriptfont2\tensy  \scriptscriptfont2\sevensy%
  \textfont\itfam=\fourteenit\def\it{\fam\itfam\fourteenit}%
  \ifprod@font
    \scriptfont\itfam=\tenit
      \scriptscriptfont\itfam=\sevenit
  \else
    \scriptfont\itfam=\fourteenit
      \scriptscriptfont\itfam=\fourteenit
  \fi
  \textfont\bffam=\fourteenbf%
    \scriptfont\bffam=\tenbf%
      \scriptscriptfont\bffam=\sevenbf%
  \def\bf{\fam\bffam\fourteenbf}%
  \textfont\slfam=\fourteensl\def\sl{\fam\slfam\fourteensl}%
  \ifprod@font
    \scriptfont\slfam=\tensl
      \scriptscriptfont\slfam=\sevensl
  \else
    \scriptfont\slfam=\fourteensl
      \scriptscriptfont\slfam=\fourteensl
  \fi
  \textfont\ttfam=\fourteentt\def\tt{\fam\ttfam\fourteentt}%
  \ifprod@font
    \scriptfont\ttfam=\tentt
      \scriptscriptfont\ttfam=\seventt
  \else
    \scriptfont\ttfam=\fourteentt
      \scriptscriptfont\ttfam=\fourteentt
  \fi
  \textfont\scfam=\fourteencsc\def\sc{\fam\scfam\fourteencsc}%
  \ifprod@font
    \scriptfont\scfam=\tencsc
      \scriptscriptfont\scfam=\sevencsc
  \else
    \scriptfont\scfam=\fourteencsc
      \scriptscriptfont\scfam=\fourteencsc
  \fi
  \textfont\sffam=\fourteensf\def\sf{\fam\sffam\fourteensf}%
  \ifprod@font
    \scriptfont\sffam=\tensf
      \scriptscriptfont\sffam=\sevensf
  \else
    \scriptfont\sffam=\fourteensf
      \scriptscriptfont\sffam=\fourteensf
  \fi
  \def\oldstyle{\fam\@ne\fourteeni}%
  \b@ls{17pt}\rm\@xivpt%
}
\def\@xivpt{}

\def\seventeenpoint{
  \def\rm{\fam0\seventeenrm}%
  \textfont0\seventeenrm  \scriptfont0\twelverm  \scriptscriptfont0\tenrm%
  \textfont1\seventeeni   \scriptfont1\twelvei   \scriptscriptfont1\teni%
  \textfont2\seventeensy  \scriptfont2\twelvesy  \scriptscriptfont2\tensy%
  \textfont\itfam=\seventeenit\def\it{\fam\itfam\seventeenit}%
  \ifprod@font
    \scriptfont\itfam=\twelveit
      \scriptscriptfont\itfam=\tenit
  \else
    \scriptfont\itfam=\seventeenit
      \scriptscriptfont\itfam=\seventeenit
  \fi
  \textfont\bffam=\seventeenbf%
    \scriptfont\bffam=\twelvebf%
      \scriptscriptfont\bffam=\tenbf%
  \def\bf{\fam\bffam\seventeenbf}%
  \textfont\slfam=\seventeensl\def\sl{\fam\slfam\seventeensl}%
  \ifprod@font
    \scriptfont\slfam=\twelvesl
      \scriptscriptfont\slfam=\tensl
  \else
    \scriptfont\slfam=\seventeensl
      \scriptscriptfont\slfam=\seventeensl
  \fi
  \textfont\ttfam=\seventeentt\def\tt{\fam\ttfam\seventeentt}%
  \ifprod@font
    \scriptfont\ttfam=\twelvett
      \scriptscriptfont\ttfam=\tentt
  \else
    \scriptfont\ttfam=\seventeentt
      \scriptscriptfont\ttfam=\seventeentt
  \fi
  \textfont\scfam=\seventeencsc\def\sc{\fam\scfam\seventeencsc}%
  \ifprod@font
    \scriptfont\scfam=\twelvecsc
      \scriptscriptfont\scfam=\tencsc
  \else
    \scriptfont\scfam=\seventeencsc
      \scriptscriptfont\scfam=\seventeencsc
  \fi
  \textfont\sffam=\seventeensf\def\sf{\fam\sffam\seventeensf}%
  \ifprod@font
    \scriptfont\sffam=\twelvesf
      \scriptscriptfont\sffam=\tensf
  \else
    \scriptfont\sffam=\seventeensf
      \scriptscriptfont\sffam=\seventeensf
  \fi
  \def\oldstyle{\fam\@ne\seventeeni}%
  \b@ls{20pt}\rm\@xviipt%
}
\def\@xviipt{}

\lineskip=1pt      \normallineskip=\lineskip
\lineskiplimit=\z@ \normallineskiplimit=\lineskiplimit


\def\loadboldmathnames{%
  \def\balpha{{\bmath{\alpha}}}%
  \def\bbeta{{\bmath{\beta}}}%
  \def\bgamma{{\bmath{\gamma}}}%
  \def\bdelta{{\bmath{\delta}}}%
  \def\bepsilon{{\bmath{\epsilon}}}%
  \def\bzeta{{\bmath{\zeta}}}%
  \def\boldeta{{\bmath{\eta}}}%
  \def\btheta{{\bmath{\theta}}}%
  \def\biota{{\bmath{\iota}}}%
  \def\bkappa{{\bmath{\kappa}}}%
  \def\blambda{{\bmath{\lambda}}}%
  \def\bmu{{\bmath{\mu}}}%
  \def\bnu{{\bmath{\nu}}}%
  \def\bxi{{\bmath{\xi}}}%
  \def\bpi{{\bmath{\pi}}}%
  \def\brho{{\bmath{\rho}}}%
  \def\bsigma{{\bmath{\sigma}}}%
  \def\btau{{\bmath{\tau}}}%
  \def\bupsilon{{\bmath{\upsilon}}}%
  \def\bphi{{\bmath{\phi}}}%
  \def\bchi{{\bmath{\chi}}}%
  \def\bpsi{{\bmath{\psi}}}%
  \def\bomega{{\bmath{\omega}}}%
  \def\bvarepsilon{{\bmath{\varepsilon}}}%
  \def\bvartheta{{\bmath{\vartheta}}}%
  \def\bvarpi{{\bmath{\varpi}}}%
  \def\bvarrho{{\bmath{\varrho}}}%
  \def\bvarsigma{{\bmath{\varsigma}}}%
  \def\bvarphi{{\bmath{\varphi}}}%
  \def\baleph{{\bmath{\aleph}}}%
  \def\bimath{{\bmath{\imath}}}%
  \def\bjmath{{\bmath{\jmath}}}%
  \def\bell{{\bmath{\ell}}}%
  \def\bwp{{\bmath{\wp}}}%
  \def\bRe{{\bmath{\Re}}}%
  \def\bIm{{\bmath{\Im}}}%
  \def\bpartial{{\bmath{\partial}}}%
  \def\binfty{{\bmath{\infty}}}%
  \def\bprime{{\bmath{\prime}}}%
  \def\bemptyset{{\bmath{\emptyset}}}%
  \def\bnabla{{\bmath{\nabla}}}%
  \def\btop{{\bmath{\top}}}%
  \def\bbot{{\bmath{\bot}}}%
  \def\btriangle{{\bmath{\triangle}}}%
  \def\bforall{{\bmath{\forall}}}%
  \def\bexists{{\bmath{\exists}}}%
  \def\bneg{{\bmath{\neg}}}%
  \def\bflat{{\bmath{\flat}}}%
  \def\bnatural{{\bmath{\natural}}}%
  \def\bsharp{{\bmath{\sharp}}}%
  \def\bclubsuit{{\bmath{\clubsuit}}}%
  \def\bdiamondsuit{{\bmath{\diamondsuit}}}%
  \def\bheartsuit{{\bmath{\heartsuit}}}%
  \def\bspadesuit{{\bmath{\spadesuit}}}%
  \def\bsmallint{{\bmath{\smallint}}}%
  \def\btriangleleft{{\bmath{\triangleleft}}}%
  \def\btriangleright{{\bmath{\triangleright}}}%
  \def\bbigtriangleup{{\bmath{\bigtriangleup}}}%
  \def\bbigtriangledown{{\bmath{\bigtriangledown}}}%
  \def\bwedge{{\bmath{\wedge}}}%
  \def\bvee{{\bmath{\vee}}}%
  \def\bcap{{\bmath{\cap}}}%
  \def\bcup{{\bmath{\cup}}}%
  \def\bddagger{{\bmath{\ddagger}}}%
  \def\bdagger{{\bmath{\dagger}}}%
  \def\bsqcap{{\bmath{\sqcap}}}%
  \def\bsqcup{{\bmath{\sqcup}}}%
  \def\buplus{{\bmath{\uplus}}}%
  \def\bamalg{{\bmath{\amalg}}}%
  \def\bdiamond{{\bmath{\diamond}}}%
  \def\bbullet{{\bmath{\bullet}}}%
  \def\bwr{{\bmath{\wr}}}%
  \def\bdiv{{\bmath{\div}}}%
  \def\bodot{{\bmath{\odot}}}%
  \def\boslash{{\bmath{\oslash}}}%
  \def\botimes{{\bmath{\otimes}}}%
  \def\bominus{{\bmath{\ominus}}}%
  \def\boplus{{\bmath{\oplus}}}%
  \def\bmp{{\bmath{\mp}}}%
  \def\bpm{{\bmath{\pm}}}%
  \def\bcirc{{\bmath{\circ}}}%
  \def\bbigcirc{{\bmath{\bigcirc}}}%
  \def\bsetminus{{\bmath{\setminus}}}%
  \def\bcdot{{\bmath{\cdot}}}%
  \def\bast{{\bmath{\ast}}}%
  \def\btimes{{\bmath{\times}}}%
  \def\bstar{{\bmath{\star}}}%
  \def\bpropto{{\bmath{\propto}}}%
  \def\bsqsubseteq{{\bmath{\sqsubseteq}}}%
  \def\bsqsupseteq{{\bmath{\sqsupseteq}}}%
  \def\bparallel{{\bmath{\parallel}}}%
  \def\bmid{{\bmath{\mid}}}%
  \def\bdashv{{\bmath{\dashv}}}%
  \def\bvdash{{\bmath{\vdash}}}%
  \def\bnearrow{{\bmath{\nearrow}}}%
  \def\bsearrow{{\bmath{\searrow}}}%
  \def\bnwarrow{{\bmath{\nwarrow}}}%
  \def\bswarrow{{\bmath{\swarrow}}}%
  \def\bLeftrightarrow{{\bmath{\Leftrightarrow}}}%
  \def\bLeftarrow{{\bmath{\Leftarrow}}}%
  \def\bRightarrow{{\bmath{\Rightarrow}}}%
  \def\bleq{{\bmath{\leq}}}%
  \def\bgeq{{\bmath{\geq}}}%
  \def\bsucc{{\bmath{\succ}}}%
  \def\bprec{{\bmath{\prec}}}%
  \def\bapprox{{\bmath{\approx}}}%
  \def\bsucceq{{\bmath{\succeq}}}%
  \def\bpreceq{{\bmath{\preceq}}}%
  \def\bsupset{{\bmath{\supset}}}%
  \def\bsubset{{\bmath{\subset}}}%
  \def\bsupseteq{{\bmath{\supseteq}}}%
  \def\bsubseteq{{\bmath{\subseteq}}}%
  \def\bin{{\bmath{\in}}}%
  \def\bni{{\bmath{\ni}}}%
  \def\bgg{{\bmath{\gg}}}%
  \def\bll{{\bmath{\ll}}}%
  \def\bnot{{\bmath{\not}}}%
  \def\bleftrightarrow{{\bmath{\leftrightarrow}}}%
  \def\bleftarrow{{\bmath{\leftarrow}}}%
  \def\brightarrow{{\bmath{\rightarrow}}}%
  \def\bmapstochar{{\bmath{\mapstochar}}}%
  \def\bsim{{\bmath{\sim}}}%
  \def\bsimeq{{\bmath{\simeq}}}%
  \def\bperp{{\bmath{\perp}}}%
  \def\bequiv{{\bmath{\equiv}}}%
  \def\basymp{{\bmath{\asymp}}}%
  \def\bsmile{{\bmath{\smile}}}%
  \def\bfrown{{\bmath{\frown}}}%
  \def\bleftharpoonup{{\bmath{\leftharpoonup}}}%
  \def\bleftharpoondown{{\bmath{\leftharpoondown}}}%
  \def\brightharpoonup{{\bmath{\rightharpoonup}}}%
  \def\brightharpoondown{{\bmath{\rightharpoondown}}}%
  \def\blhook{{\bmath{\lhook}}}%
  \def\brhook{{\bmath{\rhook}}}%
  \def\bldotp{{\bmath{\ldotp}}}%
  \def\bcdotp{{\bmath{\cdotp}}}%
}

\def\,{\relax\ifmmode \mskip\thinmuskip\else \thinspace\fi}
\let\protect=\relax

\long\def\@ifundefined#1#2#3{\expandafter\ifx\csname
  #1\endcsname\relax#2\else#3\fi}




\newtoks\math@groups \math@groups={}
\def\addtom@thgroup#1#2{#1\expandafter{\the#1#2}} 



\def\addtosizeh@ok#1#2#3#4{%
  \expandafter\def\csname @#1pt\endcsname{%
    \def\s@ze{#2}\def\ss@ze{#3}\def\sss@ze{#4}\the\math@groups%
  }%
}



\let\resetsizehook=\addtosizeh@ok


\ifprod@font
  \addtosizeh@ok{viii} {8} {6}  {5}
  \addtosizeh@ok{ix}   {9} {6}  {5}
  \addtosizeh@ok{x}    {10}{7}  {5}
  \addtosizeh@ok{xi}   {11}{8}  {6}
  \addtosizeh@ok{xiv}  {14}{10} {7}
  \addtosizeh@ok{xvii} {17}{12}{10}
\else
  \addtosizeh@ok{viii} {8}     {6}     {5}
  \addtosizeh@ok{ix}   {9}     {6}     {5}
  \addtosizeh@ok{x}    {10}    {7}     {5}
  \addtosizeh@ok{xi}   {10.95} {8}     {6}
  \addtosizeh@ok{xiv}  {14.4}  {10}    {7}
  \addtosizeh@ok{xvii} {17.28} {12}    {10}
\fi

\def\get@font#1#2#3{%
  \edef\fonts@ze{\romannumeral#3}
  \edef\fontn@me{\fonts@ze#1}
  \@ifundefined{\fontn@me}%
    {
     \global\expandafter\font\csname \fontn@me\endcsname=#2 at #3pt}%
    {}%
}

\def\ass@tfont#1#2{%
  \xdef\fam@name{\csname #1\endcsname}%
  \xdef\font@name{\csname #2\endcsname}%
  \let\textfont@name\font@name
  \textfont\fam@name\textfont@name
}

\def\ass@sfont#1#2{%
  \xdef\fam@name{\csname #1\endcsname}%
  \xdef\font@name{\csname #2\endcsname}%
  \let\textfont@name\font@name
  \scriptfont\fam@name\textfont@name
}

\def\ass@ssfont#1#2{%
  \xdef\fam@name{\csname #1\endcsname}%
  \xdef\font@name{\csname #2\endcsname}%
  \let\textfont@name\font@name
  \scriptscriptfont\fam@name\textfont@name
}


\def\NewSymbolFont#1#2{%
  \expandafter\ifx\csname sym#1fam\endcsname\relax 
    \expandafter\newfam\csname sym#1fam\endcsname
    \expandafter\edef\csname sym#1fam\endcsname{\the\allocationnumber}%
    \addtom@thgroup\math@groups{%
      \get@font{#1}{#2}{\s@ze}%
      \ass@tfont{sym#1fam}{\fontn@me}%
      \get@font{#1}{#2}{\ss@ze}%
      \ass@sfont{sym#1fam}{\fontn@me}%
      \get@font{#1}{#2}{\sss@ze}%
      \ass@ssfont{sym#1fam}{\fontn@me}%
    }%
  \else
    \errmessage{Family `#1' already defined}%
  \fi
}


\def\NewMathSymbol#1#2#3#4{%
  \edef\f@mly{\expandafter\hexnumber{\csname sym#3fam\endcsname}}%
  \mathchardef#1="#2\f@mly#4\relax
}


\newif\ifd@f

\def\NewMathDelimiter#1#2#3#4#5#6{%
  \d@ftrue
  \expandafter\ifx\csname sym#3fam\endcsname\relax
    \d@ffalse \errmessage{Family `#3' is not defined}%
  \fi
  \expandafter\ifx\csname sym#5fam\endcsname\relax
    \d@ffalse \errmessage{Family `#5' is not defined}%
  \fi
  \ifd@f
    \edef\f@mly{\expandafter\hexnumber{\csname sym#3fam\endcsname}}%
    \edef\f@mlytw@{\expandafter\hexnumber{\csname sym#5fam\endcsname}}%
    \xdef#1{\delimiter"#2\f@mly #4\f@mlytw@ #6\relax}%
  \fi
}


\def\setboxz@h{\setbox\z@\hbox}
\def\wdz@{\wd\z@}
\def\boxz@{\box\z@}
\def\setbox@ne{\setbox\@ne}
\def\wd@ne{\wd\@ne}

\def\math@atom#1#2{%
   \binrel@{#1}\binrel@@{#2}}
\def\binrel@#1{\setboxz@h{\thinmuskip0mu
  \medmuskip\m@ne mu\thickmuskip\@ne mu$#1\m@th$}%
 \setbox@ne\hbox{\thinmuskip0mu\medmuskip\m@ne mu\thickmuskip
  \@ne mu${}#1{}\m@th$}%
 \setbox\tw@\hbox{\hskip\wd@ne\hskip-\wdz@}}
\def\binrel@@#1{\ifdim\wd2<\z@\mathbin{#1}\else\ifdim\wd\tw@>\z@
 \mathrel{#1}\else{#1}\fi\fi}

\def\m@thit{1}

\def\set@skchar#1{\global\expandafter\skewchar
  \csname\fontn@me\endcsname=#1\relax}

\def\NewMathAlphabet#1#2#3{%
  \def\tst{#3}%
  \ifx\tst\empty\else 
    \expandafter\gdef\csname #1@sc\endcsname{}
  \fi
  \expandafter\def\csname #1\endcsname{
    \protect\csname @#1\endcsname}%
  \expandafter\def\csname @#1\endcsname##1{
    {%
    \begingroup
      \get@font{#1}{#2}{\s@ze}%
      \@ifundefined{#1@sc}{}{\set@skchar{#3}}%
      \ass@tfont{m@thit}{\fontn@me}%
      \get@font{#1}{#2}{\ss@ze}%
      \@ifundefined{#1@sc}{}{\set@skchar{#3}}%
      \ass@sfont{m@thit}{\fontn@me}%
      \get@font{#1}{#2}{\sss@ze}%
      \@ifundefined{#1@sc}{}{\set@skchar{#3}}%
      \ass@ssfont{m@thit}{\fontn@me}%
      \math@atom{##1}{%
      \mathchoice%
        {\hbox{$\m@th\displaystyle##1$}}%
        {\hbox{$\m@th\textstyle##1$}}%
        {\hbox{$\m@th\scriptstyle##1$}}%
        {\hbox{$\m@th\scriptscriptstyle##1$}}}%
    \endgroup
    }%
  }%
}


\newif\iffirstta  \firsttatrue

\def\set@hchar#1{\global\expandafter\hyphenchar
  \csname\fontn@me\endcsname=#1\relax}

\def\NewTextAlphabet#1#2#3{%
  \iffirstta
    \global\firsttafalse
    \newfam\scratchfam
    \edef\scrt@fam{\the\allocationnumber}%
  \fi
  \def\tst{#3}%
  \ifx\tst\empty\else 
    \expandafter\gdef\csname #1@hc\endcsname{}
  \fi
  \expandafter\def\csname #1\endcsname{
    \protect\csname t@#1\endcsname}%
  \long\expandafter\def\csname t@#1\endcsname##1{
    \ifmmode
      \typeout{Warning: do not use \expandafter\string\csname #1\endcsname
        \space in math mode}\fi%
    {%
      \get@font{#1}{#2}{\s@ze}\let\t@xtfnt=\fontn@me\relax
      \@ifundefined{#1@hc}{}{\set@hchar{#3}}%
      \ass@tfont{scrt@fam}{\fontn@me}%
      \get@font{#1}{#2}{\ss@ze}%
      \@ifundefined{#1@hc}{}{\set@hchar{#3}}%
      \ass@sfont{scrt@fam}{\fontn@me}%
      \get@font{#1}{#2}{\sss@ze}%
      \@ifundefined{#1@hc}{}{\set@hchar{#3}}%
      \ass@ssfont{scrt@fam}{\fontn@me}%
      \fam\scratchfam\csname\t@xtfnt\endcsname
    ##1%
    }%
  }%
  \expandafter\def\csname #1shape
    \endcsname{\protect\csname @#1shape\endcsname}%
  \expandafter\def\csname @#1shape\endcsname{
    \ifmmode
      \typeout{Warning: do not use \expandafter\string\csname
        #1shape\endcsname \space in math mode}\fi
      \get@font{#1}{#2}{\s@ze}\let\t@xtfnt=\fontn@me\relax
      \@ifundefined{#1@hc}{}{\set@hchar{#3}}%
      \ass@tfont{scrt@fam}{\fontn@me}%
      \get@font{#1}{#2}{\ss@ze}%
      \@ifundefined{#1@hc}{}{\set@hchar{#3}}%
      \ass@sfont{scrt@fam}{\fontn@me}%
      \get@font{#1}{#2}{\sss@ze}%
      \@ifundefined{#1@hc}{}{\set@hchar{#3}}%
      \ass@ssfont{scrt@fam}{\fontn@me}%
      \fam\scratchfam\csname\t@xtfnt\endcsname
  }%
}


\ifprod@font
  \def\math@itfnt{mtmib10}
  \def\math@syfnt{mtbsy10}
\else
  \def\math@itfnt{cmmib10}
  \def\math@syfnt{cmbsy10}
\fi

\def\m@thsy{2}

\def\bmath{\protect\@bmath}
\def\@bmath#1{%
  {%
  \begingroup
    \get@font{mthit}{\math@itfnt}{\s@ze}\set@skchar{'177}%
    \ass@tfont{m@thit}{\fontn@me}%
    \get@font{mthit}{\math@itfnt}{\ss@ze}\set@skchar{'177}%
    \ass@sfont{m@thit}{\fontn@me}%
    \get@font{mthit}{\math@itfnt}{\sss@ze}\set@skchar{'177}%
    \ass@ssfont{m@thit}{\fontn@me}%
    \get@font{mthsy}{\math@syfnt}{\s@ze}\set@skchar{'60}%
    \ass@tfont{m@thsy}{\fontn@me}%
    \get@font{mthsy}{\math@syfnt}{\ss@ze}\set@skchar{'60}%
    \ass@sfont{m@thsy}{\fontn@me}%
    \get@font{mthsy}{\math@syfnt}{\sss@ze}\set@skchar{'60}%
    \ass@ssfont{m@thsy}{\fontn@me}%
    \math@atom{#1}{%
    \mathchoice%
      {\hbox{$\m@th\displaystyle#1$}}%
      {\hbox{$\m@th\textstyle#1$}}%
      {\hbox{$\m@th\scriptstyle#1$}}%
      {\hbox{$\m@th\scriptscriptstyle#1$}}}%
  \endgroup
  }%
}



\def\diameter{{\ifmmode\mathchoice
{\ooalign{\hfil\hbox{$\displaystyle/$}\hfil\crcr
{\hbox{$\displaystyle\mathchar"20D$}}}}
{\ooalign{\hfil\hbox{$\textstyle/$}\hfil\crcr
{\hbox{$\textstyle\mathchar"20D$}}}}
{\ooalign{\hfil\hbox{$\scriptstyle/$}\hfil\crcr
{\hbox{$\scriptstyle\mathchar"20D$}}}}
{\ooalign{\hfil\hbox{$\scriptscriptstyle/$}\hfil\crcr
{\hbox{$\scriptscriptstyle\mathchar"20D$}}}}
\else{\ooalign{\hfil/\hfil\crcr\mathhexbox20D}}%
\fi}}

\def\sq{\ifmmode\squareforqed\else{\unskip\nobreak\hfil
\penalty50\hskip1em\null\nobreak\hfil\squareforqed
\parfillskip=0pt\finalhyphendemerits=0\endgraf}\fi}
\def\squareforqed{\hbox{\rlap{$\sqcap$}$\sqcup$}}


\def\bbbc{{\mathchoice {\setbox0=\hbox{$\displaystyle\rm C$}\hbox{\hbox
to0pt{\kern0.4\wd0\vrule height0.9\ht0\hss}\box0}}
{\setbox0=\hbox{$\textstyle\rm C$}\hbox{\hbox
to0pt{\kern0.4\wd0\vrule height0.9\ht0\hss}\box0}}
{\setbox0=\hbox{$\scriptstyle\rm C$}\hbox{\hbox
to0pt{\kern0.4\wd0\vrule height0.9\ht0\hss}\box0}}
{\setbox0=\hbox{$\scriptscriptstyle\rm C$}\hbox{\hbox
to0pt{\kern0.4\wd0\vrule height0.9\ht0\hss}\box0}}}}
\def\bbbq{{\mathchoice {\setbox0=\hbox{$\displaystyle\rm
Q$}\hbox{\raise
0.15\ht0\hbox to0pt{\kern0.4\wd0\vrule height0.8\ht0\hss}\box0}}
{\setbox0=\hbox{$\textstyle\rm Q$}\hbox{\raise
0.15\ht0\hbox to0pt{\kern0.4\wd0\vrule height0.8\ht0\hss}\box0}}
{\setbox0=\hbox{$\scriptstyle\rm Q$}\hbox{\raise
0.15\ht0\hbox to0pt{\kern0.4\wd0\vrule height0.7\ht0\hss}\box0}}
{\setbox0=\hbox{$\scriptscriptstyle\rm Q$}\hbox{\raise
0.15\ht0\hbox to0pt{\kern0.4\wd0\vrule height0.7\ht0\hss}\box0}}}}
\def\bbbt{{\mathchoice {\setbox0=\hbox{$\displaystyle\rm
T$}\hbox{\hbox to0pt{\kern0.3\wd0\vrule height0.9\ht0\hss}\box0}}
{\setbox0=\hbox{$\textstyle\rm T$}\hbox{\hbox
to0pt{\kern0.3\wd0\vrule height0.9\ht0\hss}\box0}}
{\setbox0=\hbox{$\scriptstyle\rm T$}\hbox{\hbox
to0pt{\kern0.3\wd0\vrule height0.9\ht0\hss}\box0}}
{\setbox0=\hbox{$\scriptscriptstyle\rm T$}\hbox{\hbox
to0pt{\kern0.3\wd0\vrule height0.9\ht0\hss}\box0}}}}
\def\bbbs{{\mathchoice
{\setbox0=\hbox{$\displaystyle     \rm S$}\hbox{\raise0.5\ht0\hbox
to0pt{\kern0.35\wd0\vrule height0.45\ht0\hss}\hbox
to0pt{\kern0.55\wd0\vrule height0.5\ht0\hss}\box0}}
{\setbox0=\hbox{$\textstyle        \rm S$}\hbox{\raise0.5\ht0\hbox
to0pt{\kern0.35\wd0\vrule height0.45\ht0\hss}\hbox
to0pt{\kern0.55\wd0\vrule height0.5\ht0\hss}\box0}}
{\setbox0=\hbox{$\scriptstyle      \rm S$}\hbox{\raise0.5\ht0\hbox
to0pt{\kern0.35\wd0\vrule height0.45\ht0\hss}\raise0.05\ht0\hbox
to0pt{\kern0.5\wd0\vrule height0.45\ht0\hss}\box0}}
{\setbox0=\hbox{$\scriptscriptstyle\rm S$}\hbox{\raise0.5\ht0\hbox
to0pt{\kern0.4\wd0\vrule height0.45\ht0\hss}\raise0.05\ht0\hbox
to0pt{\kern0.55\wd0\vrule height0.45\ht0\hss}\box0}}}}
\def\bbbz{{\mathchoice {\hbox{$\sf\textstyle Z\kern-0.4em Z$}}
{\hbox{$\sf\textstyle Z\kern-0.4em Z$}}
{\hbox{$\sf\scriptstyle Z\kern-0.3em Z$}}
{\hbox{$\sf\scriptscriptstyle Z\kern-0.2em Z$}}}}


\def\Nulle{0} 
\def\Afe{1}   
\def\Hae{2}   
\def\Hbe{3}   
\def\Hce{4}   
\def\Hde{5}   


\newcount\LastMac       \LastMac=\Nulle

\newskip\half      \half=5.5pt plus 1.5pt minus 2.25pt
\newskip\one       \one=11pt plus 3pt minus 5.5pt
\newskip\onehalf   \onehalf=16.5pt plus 5.5pt minus 8.25pt
\newskip\two       \two=22pt plus 5.5pt minus 11pt

\def\Half{\addvspace{\half}}
\def\One{\addvspace{\one}}
\def\OneHalf{\addvspace{\onehalf}}
\def\Two{\addvspace{\two}}

\def\Raggedright{
  \rightskip=\z@ plus \hsize\relax
}

\def\Fullout{
  \rightskip=\z@\relax
}

\def\Hang#1#2{
  \hangindent=#1%
  \hangafter=#2\relax
}


\newif\ifsp@page
\def\pagestyle#1{\csname ps@#1\endcsname}
\def\thispagestyle#1{\global\sp@pagetrue\gdef\sp@type{#1}}

\def\ps@titlepage{%
  \def\@oddhead{\eightpoint\noindent \the\CatchLine
    \ifprod@font\else\qquad Printed\ \today\qquad
      (MN plain \TeX\ macros\ v\@version)\fi \hfil}%
  \let\@evenhead=\@oddhead
  \def\@oddfoot{\eightpoint\copyright\ \@pubyear\ RAS\hfil}%
  \def\@evenfoot{\hfil\eightpoint\noindent\copyright\ \@pubyear\ RAS}%
}

\def\ps@headings{%
  \def\@oddhead{\elevenpoint\it\noindent
    \hfill\the\RightHeader\hskip1.5em\rm\folio}%
  \def\@evenhead{\elevenpoint\noindent
    \folio\hskip1.5em\it\the\LeftHeader\hfill}%
  \def\@oddfoot{\eightpoint\noindent\copyright\ \@pubyear\ RAS,
    MNRAS {\bf \@volume}, \@pagerange\hfil}%
  \def\@evenfoot{\hfil\eightpoint\copyright\ \@pubyear\ RAS,
    MNRAS {\bf \@volume}, \@pagerange}%
}

\def\ps@plate{%
  \def\@oddhead{\eightpoint\noindent\plt@cap\hfil}%
  \def\@evenhead{\eightpoint\noindent\plt@cap\hfil}%
  \def\@oddfoot{\eightpoint\noindent\copyright\ \@pubyear\ RAS,
    MNRAS {\bf \@volume}, \@pagerange\hfil}%
  \def\@evenfoot{\hfil\eightpoint\copyright\ \@pubyear\ RAS,
    MNRAS {\bf \@volume}, \@pagerange}%
}



\def\title#1{
  \bgroup
    \vbox to 8pt{\vss}%
    \seventeenpoint
    \Raggedright
    \noindent \strut{\bf #1}\par
  \egroup
}

\def\author#1{
  \bgroup
    \ifnum\LastMac=\Afe \OneHalf\else \vskip 21pt\fi
    \fourteenpoint
    \Raggedright
    \noindent \strut #1\par
    \vskip 3pt%
  \egroup
}

\def\affiliation#1{
  \bgroup
    \vskip -4pt%
    \eightpoint
    \Raggedright
    \noindent \strut {\it #1}\par
  \egroup
  \LastMac=\Afe\relax
}

\def\acceptedline#1{
  \bgroup
    \Two
    \eightpoint
    \Raggedright
    \noindent \strut #1\par
  \egroup
}

\long\def\abstract#1{%
  \bgroup
    \vskip 20pt%
    \leftskip 11pc\rightskip\z@
    \noindent{\ninebf ABSTRACT}\par
    \tenpoint
    \Fullout
    \noindent #1\par
  \egroup
}

\long\def\keywords#1{
  \bgroup
    \Half
    \leftskip 11pc\rightskip\z@
    \tenpoint
    \Fullout
    \noindent\hbox{\bf Key words:}\ #1\par
  \egroup
}


\def\maketitle{%
  \EndOpening
  \ifsinglecol \else \MakePage\fi
}


\def\pageoffset#1#2{\hoffset=#1\relax\voffset=#2\relax}


\def\@nameuse#1{\csname #1\endcsname}
\def\arabic#1{\@arabic{\@nameuse{#1}}}
\def\alph#1{\@alph{\@nameuse{#1}}}
\def\Alph#1{\@Alph{\@nameuse{#1}}}
\def\@arabic#1{\number #1}
\def\@Alph#1{\ifcase#1\or A\or B\or C\or D\else\@Ialph{#1}\fi}
\def\@Ialph#1{\ifcase#1\or \or \or \or \or E\or F\or G\or H\or I\or J\or
   K\or L\or M\or N\or O\or P\or Q\or R\or S\or T\or U\or V\or W\or X\or
   Y\or Z\else\errmessage{Counter out of range}\fi}
\def\@alph#1{\ifcase#1\or a\or b\or c\or d\else\@ialph{#1}\fi}
\def\@ialph#1{\ifcase#1\or \or \or \or \or e\or f\or g\or h\or i\or j\or
   k\or l\or m\or n\or o\or p\or q\or r\or s\or t\or u\or v\or w\or x\or y\or
   z\else\errmessage{Counter out of range}\fi}


\newcount\Eqnno
\newcount\SubEqnno

\def\theeq{\arabic{Eqnno}}
\def\thesubeq{\alph{SubEqnno}}

\def\stepeq{\relax
  \global\SubEqnno \z@
  \global\advance\Eqnno \@ne\relax
  {\rm (\theeq)}%
}

\def\startsubeq{\relax
  \global\SubEqnno \z@
  \global\advance\Eqnno \@ne\relax
  \stepsubeq
}

\def\stepsubeq{\relax
  \global\advance\SubEqnno \@ne\relax
  {\rm (\theeq\thesubeq)}%
}


\newcount\Sec        
\newcount\SecSec
\newcount\SecSecSec

\def\thesection{\arabic{Sec}}
\def\thesubsection{\thesection.\arabic{SecSec}}
\def\thesubsubsection{\thesubsection.\arabic{SecSecSec}}

\Sec=\z@

\def\:{\let\@sptoken= } \:  
\def\:{\@xifnch} \expandafter\def\: {\futurelet\@tempc\@ifnch}

\def\@ifnextchar#1#2#3{%
  \let\@tempMACe #1%
  \def\@tempMACa{#2}%
  \def\@tempMACb{#3}%
  \futurelet \@tempMACc\@ifnch%
}

\def\@ifnch{%
\ifx \@tempMACc \@sptoken%
  \let\@tempMACd\@xifnch%
\else%
  \ifx \@tempMACc \@tempMACe%
    \let\@tempMACd\@tempMACa%
  \else%
    \let\@tempMACd\@tempMACb%
  \fi%
\fi%
\@tempMACd%
}

\def\@ifstar#1#2{\@ifnextchar *{\def\@tempMACa*{#1}\@tempMACa}{#2}}

\newskip\@tempskipb

\def\addvspace#1{%
  \ifvmode\else \endgraf\fi%
  \ifdim\lastskip=\z@%
    \vskip #1\relax%
  \else%
    \@tempskipb#1\relax\@xaddvskip%
  \fi%
}

\def\@xaddvskip{%
  \ifdim\lastskip<\@tempskipb%
    \vskip-\lastskip%
    \vskip\@tempskipb\relax%
  \else%
    \ifdim\@tempskipb<\z@%
      \ifdim\lastskip<\z@ \else%
        \advance\@tempskipb\lastskip%
        \vskip-\lastskip\vskip\@tempskipb%
      \fi%
    \fi%
  \fi%
}

\newskip\@tmpSKIP

\def\addpen#1{%
  \ifvmode
    \if@nobreak
    \else
      \ifdim\lastskip=\z@
        \penalty#1\relax
      \else
        \@tmpSKIP=\lastskip
        \vskip -\lastskip
        \penalty#1\vskip\@tmpSKIP
      \fi
    \fi
  \fi
}

\newcount\@clubpen   \@clubpen=\clubpenalty
\newif\if@nobreak    \@nobreakfalse

\def\@noafterindent{%
  \global\@nobreaktrue
  \everypar{\if@nobreak
              \global\@nobreakfalse
              \clubpenalty \@M
              {\setbox\z@\lastbox}%
              \LastMac=\Nulle\relax%
            \else
              \clubpenalty \@clubpen
              \everypar{}%
            \fi}%
}

\newcount\gds@cbrk   \gds@cbrk=-300

\def\@nohdbrk{\interlinepenalty \@M\relax}

\let\@par=\par
\def\@restorepar{\def\par{\@par}}

\newif\if@endpe   \@endpefalse
 
\def\@doendpe{\@endpetrue \@nobreakfalse \LastMac=\Nulle\relax%
     \def\par{\@restorepar\everypar{}\par\@endpefalse}%
              \everypar{\setbox\z@\lastbox\everypar{}\@endpefalse}%
}

\def\section{\@ifstar{\@ssection}{\@section}}

\def\@section#1{
  \if@nobreak
    \everypar{}%
    \ifnum\LastMac=\Hae \addvspace{\half}\fi
  \else
    \addpen{\gds@cbrk}%
    \addvspace{\two}%
  \fi
  \bgroup
    \ninepoint\bf
    \Raggedright
    \global\advance\Sec \@ne
    \ifappendix
      \global\Eqnno=\z@ \global\SubEqnno=\z@\relax
      \def\ch@ck{#1}%
      \ifx\ch@ck\empty \def\c@lon{}\else\def\c@lon{:}\fi
      \noindent\@nohdbrk APPENDIX\ \thesection\c@lon\hskip 0.5em%
        \uppercase{#1}\par
    \else
      \noindent\@nohdbrk\thesection\hskip 1pc \uppercase{#1}\par
    \fi
    \global\SecSec=\z@
  \egroup
  \nobreak
  \vskip\half
  \nobreak
  \@noafterindent
  \LastMac=\Hae\relax
}

\def\@ssection#1{
  \if@nobreak
    \everypar{}%
    \ifnum\LastMac=\Hae \addvspace{\half}\fi
  \else
    \addpen{\gds@cbrk}%
    \addvspace{\two}%
  \fi
  \bgroup
    \ninepoint\bf
    \Raggedright
    \noindent\@nohdbrk\uppercase{#1}\par
  \egroup
  \nobreak
  \vskip\half
  \nobreak
  \@noafterindent
  \LastMac=\Hae\relax
}

\def\subsection{\@ifstar{\@ssubsection}{\@subsection}}

\def\@subsection#1{
  \if@nobreak
    \everypar{}%
    \ifnum\LastMac=\Hae \addvspace{1pt plus 1pt minus .5pt}\fi
  \else
    \addpen{\gds@cbrk}%
    \addvspace{\onehalf}%
  \fi
  \bgroup
    \ninepoint\bf
    \Raggedright
    \global\advance\SecSec \@ne
    \noindent\@nohdbrk\thesubsection \hskip 1pc\relax #1\par
    \global\SecSecSec=\z@
  \egroup
  \nobreak
  \vskip\half
  \nobreak
  \@noafterindent
  \LastMac=\Hbe\relax
}

\def\@ssubsection#1{
  \if@nobreak
    \everypar{}%
    \ifnum\LastMac=\Hae \addvspace{1pt plus 1pt minus .5pt}\fi
  \else
    \addpen{\gds@cbrk}%
    \addvspace{\onehalf}%
  \fi
  \bgroup
    \ninepoint\bf
    \Raggedright
    \noindent\@nohdbrk #1\par
  \egroup
  \nobreak
  \vskip\half
  \nobreak
  \@noafterindent
  \LastMac=\Hbe\relax
}

\def\subsubsection{\@ifstar{\@ssubsubsection}{\@subsubsection}}

\def\@subsubsection#1{
  \if@nobreak
    \everypar{}%
    \ifnum\LastMac=\Hbe \addvspace{1pt plus 1pt minus .5pt}\fi
  \else
    \addpen{\gds@cbrk}%
    \addvspace{\onehalf}%
  \fi
  \bgroup
    \ninepoint\it
    \Raggedright
    \global\advance\SecSecSec \@ne
    \noindent\@nohdbrk\thesubsubsection \hskip 1pc\relax #1\par
  \egroup
  \nobreak
  \vskip\half
  \nobreak
  \@noafterindent
  \LastMac=\Hce\relax
}

\def\@ssubsubsection#1{
  \if@nobreak
    \everypar{}%
    \ifnum\LastMac=\Hbe \addvspace{1pt plus 1pt minus .5pt}\fi
  \else
    \addpen{\gds@cbrk}%
    \addvspace{\onehalf}%
  \fi
  \bgroup
    \ninepoint\it
    \Raggedright
    \noindent\@nohdbrk #1\par
  \egroup
  \nobreak
  \vskip\half
  \nobreak
  \@noafterindent
  \LastMac=\Hce\relax
}

\def\paragraph#1{
  \if@nobreak
    \everypar{}%
  \else
    \addpen{\gds@cbrk}%
    \addvspace{\one}%
  \fi%
  \bgroup%
    \ninepoint\it
    \noindent #1\ \nobreak%
  \egroup
  \LastMac=\Hde\relax
  \ignorespaces
}


\newif\ifappendix

\def\appendix{%
  \global\appendixtrue
  \def\thesection{\Alph{Sec}}%
  \def\thesubsection{\thesection\arabic{SecSec}}%
  \def\theeq{\thesection\arabic{Eqnno}}%
  \Sec=\z@ \SecSec=\z@ \SecSecSec=\z@ \Eqnno=\z@ \SubEqnno=\z@\relax
}


\let\tx=\relax 


\def\beginlist{%
  \par\if@nobreak \else\addvspace{\half}\fi%
  \bgroup%
    \ninepoint
    \let\item=\list@item%
}

\def\list@item{%
  \par\noindent\hskip 1em\relax%
  \ignorespaces%
}

\def\endlist{\par\egroup\addvspace{\half}\@doendpe}


\def\beginrefs{%
  \par
  \bgroup
    \eightpoint
    \Fullout
    \let\bibitem=\bib@item
}

\def\bib@item{%
  \par\parindent=1.5em\Hang{1.5em}{1}%
  \everypar={\Hang{1.5em}{1}\ignorespaces}%
  \noindent\ignorespaces
}

\def\endrefs{\par\egroup\@doendpe}


\newtoks\CatchLine

\def\@journal{Mon.\ Not.\ R.\ Astron.\ Soc.\ }  
\def\@pubyear{1994}        
\def\@pagerange{000--000}  
\def\@volume{000}          
\def\@microfiche{}         %

\def\pubyear#1{\gdef\@pubyear{#1}\@makecatchline}
\def\pagerange#1{\gdef\@pagerange{#1}\@makecatchline}
\def\volume#1{\gdef\@volume{#1}\@makecatchline}
\def\microfiche#1{\gdef\@microfiche{and Microfiche\ #1}\@makecatchline}

\def\@makecatchline{%
  \global\CatchLine{%
    {\rm \@journal {\bf \@volume},\ \@pagerange\ (\@pubyear)\ \@microfiche}}%
}

\@makecatchline 

\newtoks\LeftHeader
\def\shortauthor#1{
  \global\LeftHeader{#1}%
}

\newtoks\RightHeader
\def\shorttitle#1{
  \global\RightHeader{#1}%
}

\def\PageHead{
  \begingroup
    \ifsp@page
      \csname ps@\sp@type\endcsname
    \fi
    \ifodd\pageno
      \let\the@head=\@oddhead
    \else
      \let\the@head=\@evenhead
    \fi
    \vbox to \z@{\vskip-22.5\p@%
      \hbox to \PageWidth{\vbox to8.5\p@{}%
        \the@head
      }%
    \vss}%
  \endgroup
  \nointerlineskip
}

\gdef\PageFoot{%
  \nointerlineskip%
  \begingroup
  \ifsp@page
    \csname ps@\sp@type\endcsname
    \global\sp@pagefalse
  \fi
  \vbox to 22pt{\vfil%
    \hbox to \PageWidth{%
      \eightpoint\strut\noindent
      \ifodd\pageno
        \@oddfoot
      \else
        \@evenfoot
      \fi
    }%
  }%
  \endgroup
}

\def\today{%
  \number\day\space
  \ifcase\month\or January\or February\or March\or April\or May\or June\or
    July\or August\or September\or October\or November\or December\fi
  \space\number\year%
}

\def\authorcomment#1{%
  \gdef\PageFoot{%
    \nointerlineskip%
    \vbox to 20pt{\vfil%
      \hbox to \PageWidth{\elevenpoint\noindent \hfil #1 \hfil}}%
  }%
}


\newif\ifplate@page
\newbox\plt@box

\def\beginplatepage{%
  \let\plate=\plate@head
  \let\caption=\fig@caption
  \global\setbox\plt@box=\vbox\bgroup
  \TEMPDIMEN=\PageWidth 
  \hsize=\PageWidth\relax
}

\def\endplatepage{\par\egroup\global\plate@pagetrue}
\def\plate@head#1{\gdef\plt@cap{#1}}


\def\letters{%
  \gdef\folio{\ifnum\pageno<\z@ L\romannumeral-\pageno
    \else L\number\pageno \fi}%
}


\newdimen\mathindent

\global\mathindent=\z@
\global\everydisplay{\global\@dspwd=\displaywidth\displaysetup}


\def\@displaylines#1{
  {}$\displ@y\hbox{\vbox{\halign{$\@lign\hfil\displaystyle##\hfil$\crcr
  #1\crcr}}}${}%
}

\def\@eqalign#1{\null\vcenter{\openup\jot\m@th
  \ialign{\strut\hfil$\displaystyle{##}$&$\displaystyle{{}##}$\hfil
      \crcr#1\crcr}}%
}

\def\@eqalignno#1{
  \global\advance\@dspwd by -\mathindent%
  {}$\displ@y\hbox{\vbox{\halign to\@dspwd%
  {\hfil$\@lign\displaystyle{##}$\tabskip\z@skip
  &$\@lign\displaystyle{{}##}$\hfil\tabskip\centering
  &\llap{$\@lign##$}\tabskip\z@skip\crcr
  #1\crcr}}}${}%
}


\global\let\displaylines=\@displaylines
\global\let\eqalign=\@eqalign
\global\let\eqalignno=\@eqalignno
\global\let\leqalignno=\@eqalignno

\newdimen\@dspwd   \@dspwd=\z@
\newif\if@eqno
\newif\if@leqno
\newtoks\@eqn
\newtoks\@eq

\def\displaysetup#1$${\displaytest#1\eqno\eqno\displaytest}

\def\displaytest#1\eqno#2\eqno#3\displaytest{%
 \if!#3!\ldisplaytest#1\leqno\leqno\ldisplaytest
 \else\@eqnotrue\@leqnofalse\@eqn={#2}\@eq={#1}\fi
 \generaldisplay$$}

\def\ldisplaytest#1\leqno#2\leqno#3\ldisplaytest{%
\@eq={#1}%
 \if!#3!\@eqnofalse\else\@eqnotrue\@leqnotrue
  \@eqn={#2}\fi}

\def\generaldisplay{%
  \if@eqno
    \if@leqno
      \hbox to \displaywidth{\noindent
        \rlap{$\displaystyle\the\@eqn$}%
        \hskip\mathindent$\displaystyle\the\@eq$\hfil}%
    \else
      \hbox to \displaywidth{\noindent
        \hskip\mathindent
        $\displaystyle\the\@eq$\hfil$\displaystyle\the\@eqn$}%
    \fi
  \else
    \hbox to \displaywidth{\noindent
      \hskip\mathindent$\displaystyle\the\@eq$\hfil}%
  \fi
}


\def\@notice{%
  \par\Two%
  \noindent{\b@ls{11pt}\ninerm This paper has been produced using the
    Royal Astronomical Society/Blackwell Science \TeX\ macros.\par}%
}

\outer\def\bye{\@notice\par\vfill\supereject\end}


\def\start@mess{%
  Monthly notices of the RAS journal style (\@typeface)\space
    v\@version,\space \@verdate.%
}

\everyjob{\Warn{\start@mess}}



\newif\if@debug \@debugfalse  

\def\Print#1{\if@debug\immediate\write16{#1}\else \fi}
\def\Warn#1{\immediate\write16{#1}}
\def\wlog#1{}

\newcount\Iteration 

\def\Single{0} \def\Double{1}                 
\def\Figure{0} \def\Table{1}                  

\def\InStack{0}  
\def\InZoneA{1}
\def\InZoneB{2}
\def\InZoneC{3}

\newcount\TEMPCOUNT 
\newdimen\TEMPDIMEN 
\newbox\TEMPBOX     
\newbox\VOIDBOX     

\newcount\LengthOfStack 
\newcount\MaxItems      
\newcount\StackPointer
\newcount\Point         
\newcount\NextFigure    
\newcount\NextTable     
\newcount\NextItem      

\newcount\StatusStack   
\newcount\NumStack      
\newcount\TypeStack     
\newcount\SpanStack     
\newcount\BoxStack      

\newcount\ItemSTATUS    
\newcount\ItemNUMBER    
\newcount\ItemTYPE      
\newcount\ItemSPAN      
\newbox\ItemBOX         
\newdimen\ItemSIZE      

\newdimen\PageHeight    
\newdimen\TextLeading   
\newdimen\Feathering    
\newcount\LinesPerPage  
\newdimen\ColumnWidth   
\newdimen\ColumnGap     
\newdimen\PageWidth     
\newdimen\BodgeHeight   
\newcount\Leading       

\newdimen\ZoneBSize  
\newdimen\TextSize   
\newbox\ZoneABOX     
\newbox\ZoneBBOX     
\newbox\ZoneCBOX     

\newif\ifFirstSingleItem
\newif\ifFirstZoneA
\newif\ifMakePageInComplete
\newif\ifMoreFigures \MoreFiguresfalse 
\newif\ifMoreTables  \MoreTablesfalse  

\newif\ifFigInZoneB 
\newif\ifFigInZoneC 
\newif\ifTabInZoneB 
\newif\ifTabInZoneC

\newif\ifZoneAFullPage

\newbox\MidBOX    
\newbox\LeftBOX
\newbox\RightBOX
\newbox\PageBOX   

\newif\ifLeftCOL  
\LeftCOLtrue

\newdimen\ZoneBAdjust

\newcount\ItemFits
\def\Yes{1}
\def\No{2}


\MaxItems=15
\NextFigure=\z@        
\NextTable=\@ne

\BodgeHeight=6pt
\TextLeading=11pt    
\Leading=11
\Feathering=\z@      
\LinesPerPage=61     
\topskip=\TextLeading
\ColumnWidth=20pc    
\ColumnGap=2pc       

\newskip\ItemSepamount  
\ItemSepamount=\TextLeading plus \TextLeading minus 4pt

\parskip=\z@ plus .1pt
\parindent=18pt
\widowpenalty=\z@
\clubpenalty=10000
\tolerance=1500
\hbadness=1500
\abovedisplayskip=6pt plus 2pt minus 1pt
\belowdisplayskip=6pt plus 2pt minus 1pt
\abovedisplayshortskip=6pt plus 2pt minus 1pt
\belowdisplayshortskip=6pt plus 2pt minus 1pt

\frenchspacing

\ninepoint 

\PageHeight=682pt
\PageWidth=2\ColumnWidth
\advance\PageWidth by \ColumnGap

\pagestyle{headings}




\newcount\DUMMY \StatusStack=\allocationnumber
\newcount\DUMMY \newcount\DUMMY \newcount\DUMMY 
\newcount\DUMMY \newcount\DUMMY \newcount\DUMMY 
\newcount\DUMMY \newcount\DUMMY \newcount\DUMMY
\newcount\DUMMY \newcount\DUMMY \newcount\DUMMY 
\newcount\DUMMY \newcount\DUMMY \newcount\DUMMY

\newcount\DUMMY \NumStack=\allocationnumber
\newcount\DUMMY \newcount\DUMMY \newcount\DUMMY 
\newcount\DUMMY \newcount\DUMMY \newcount\DUMMY 
\newcount\DUMMY \newcount\DUMMY \newcount\DUMMY 
\newcount\DUMMY \newcount\DUMMY \newcount\DUMMY 
\newcount\DUMMY \newcount\DUMMY \newcount\DUMMY

\newcount\DUMMY \TypeStack=\allocationnumber
\newcount\DUMMY \newcount\DUMMY \newcount\DUMMY 
\newcount\DUMMY \newcount\DUMMY \newcount\DUMMY 
\newcount\DUMMY \newcount\DUMMY \newcount\DUMMY 
\newcount\DUMMY \newcount\DUMMY \newcount\DUMMY 
\newcount\DUMMY \newcount\DUMMY \newcount\DUMMY

\newcount\DUMMY \SpanStack=\allocationnumber
\newcount\DUMMY \newcount\DUMMY \newcount\DUMMY 
\newcount\DUMMY \newcount\DUMMY \newcount\DUMMY 
\newcount\DUMMY \newcount\DUMMY \newcount\DUMMY 
\newcount\DUMMY \newcount\DUMMY \newcount\DUMMY 
\newcount\DUMMY \newcount\DUMMY \newcount\DUMMY

\newbox\DUMMY   \BoxStack=\allocationnumber
\newbox\DUMMY   \newbox\DUMMY \newbox\DUMMY 
\newbox\DUMMY   \newbox\DUMMY \newbox\DUMMY 
\newbox\DUMMY   \newbox\DUMMY \newbox\DUMMY 
\newbox\DUMMY   \newbox\DUMMY \newbox\DUMMY 
\newbox\DUMMY   \newbox\DUMMY \newbox\DUMMY

\def\wlog{\immediate\write\m@ne}


\def\GetItemAll#1{%
 \GetItemSTATUS{#1}
 \GetItemNUMBER{#1}
 \GetItemTYPE{#1}
 \GetItemSPAN{#1}
 \GetItemBOX{#1}
}

\def\GetItemSTATUS#1{%
 \Point=\StatusStack
 \advance\Point by #1
 \global\ItemSTATUS=\count\Point
}

\def\GetItemNUMBER#1{%
 \Point=\NumStack
 \advance\Point by #1
 \global\ItemNUMBER=\count\Point
}

\def\GetItemTYPE#1{%
 \Point=\TypeStack
 \advance\Point by #1
 \global\ItemTYPE=\count\Point
}

\def\GetItemSPAN#1{%
 \Point\SpanStack
 \advance\Point by #1
 \global\ItemSPAN=\count\Point
}

\def\GetItemBOX#1{%
 \Point=\BoxStack
 \advance\Point by #1
 \global\setbox\ItemBOX=\vbox{\copy\Point}
 \global\ItemSIZE=\ht\ItemBOX
 \global\advance\ItemSIZE by \dp\ItemBOX
 \TEMPCOUNT=\ItemSIZE
 \divide\TEMPCOUNT by \Leading
 \divide\TEMPCOUNT by 65536
 \advance\TEMPCOUNT \@ne
 \ItemSIZE=\TEMPCOUNT pt
 \global\multiply\ItemSIZE by \Leading
}


\def\JoinStack{%
 \ifnum\LengthOfStack=\MaxItems 
  \Warn{WARNING: Stack is full...some items will be lost!}
 \else
  \Point=\StatusStack
  \advance\Point by \LengthOfStack
  \global\count\Point=\ItemSTATUS
  \Point=\NumStack
  \advance\Point by \LengthOfStack
  \global\count\Point=\ItemNUMBER
  \Point=\TypeStack
  \advance\Point by \LengthOfStack
  \global\count\Point=\ItemTYPE
  \Point\SpanStack
  \advance\Point by \LengthOfStack
  \global\count\Point=\ItemSPAN
  \Point=\BoxStack
  \advance\Point by \LengthOfStack
  \global\setbox\Point=\vbox{\copy\ItemBOX}
  \global\advance\LengthOfStack \@ne
  \ifnum\ItemTYPE=\Figure 
   \global\MoreFigurestrue
  \else
   \global\MoreTablestrue
  \fi
 \fi
}


\def\LeaveStack#1{%
 {\Iteration=#1
 \loop
 \ifnum\Iteration<\LengthOfStack
  \advance\Iteration \@ne
  \GetItemSTATUS{\Iteration}
   \advance\Point by \m@ne
   \global\count\Point=\ItemSTATUS
  \GetItemNUMBER{\Iteration}
   \advance\Point by \m@ne
   \global\count\Point=\ItemNUMBER
  \GetItemTYPE{\Iteration}
   \advance\Point by \m@ne
   \global\count\Point=\ItemTYPE
  \GetItemSPAN{\Iteration}
   \advance\Point by \m@ne
   \global\count\Point=\ItemSPAN
  \GetItemBOX{\Iteration}
   \advance\Point by \m@ne
   \global\setbox\Point=\vbox{\copy\ItemBOX}
 \repeat}
 \global\advance\LengthOfStack by \m@ne
}


\newif\ifStackNotClean

\def\CleanStack{%
 \StackNotCleantrue
 {\Iteration=\z@
  \loop
   \ifStackNotClean
    \GetItemSTATUS{\Iteration}
    \ifnum\ItemSTATUS=\InStack
     \advance\Iteration \@ne
     \else
      \LeaveStack{\Iteration}
    \fi
   \ifnum\LengthOfStack<\Iteration
    \StackNotCleanfalse
   \fi
 \repeat}
}


\def\FindItem#1#2{%
 \global\StackPointer=\m@ne 
 {\Iteration=\z@
  \loop
  \ifnum\Iteration<\LengthOfStack
   \GetItemSTATUS{\Iteration}
   \ifnum\ItemSTATUS=\InStack
    \GetItemTYPE{\Iteration}
    \ifnum\ItemTYPE=#1
     \GetItemNUMBER{\Iteration}
     \ifnum\ItemNUMBER=#2
      \global\StackPointer=\Iteration
      \Iteration=\LengthOfStack 
     \fi
    \fi
   \fi
  \advance\Iteration \@ne
 \repeat}
}


\def\FindNext{%
 \global\StackPointer=\m@ne 
 {\Iteration=\z@
  \loop
  \ifnum\Iteration<\LengthOfStack
   \GetItemSTATUS{\Iteration}
   \ifnum\ItemSTATUS=\InStack
    \GetItemTYPE{\Iteration}
   \ifnum\ItemTYPE=\Figure
    \ifMoreFigures
      \global\NextItem=\Figure
      \global\StackPointer=\Iteration
      \Iteration=\LengthOfStack 
    \fi
   \fi
   \ifnum\ItemTYPE=\Table
    \ifMoreTables
      \global\NextItem=\Table
      \global\StackPointer=\Iteration
      \Iteration=\LengthOfStack 
    \fi
   \fi
  \fi
  \advance\Iteration \@ne
 \repeat}
}


\def\ChangeStatus#1#2{%
 \Point=\StatusStack
 \advance\Point by #1
 \global\count\Point=#2
}



\def\Zone{\InZoneA}

\ZoneBAdjust=\z@

\def\MakePage{
 \global\ZoneBSize=\PageHeight
 \global\TextSize=\ZoneBSize
 \global\ZoneAFullPagefalse
 \global\topskip=\TextLeading
 \MakePageInCompletetrue
 \MoreFigurestrue
 \MoreTablestrue
 \FigInZoneBfalse
 \FigInZoneCfalse
 \TabInZoneBfalse
 \TabInZoneCfalse
 \global\FirstSingleItemtrue
 \global\FirstZoneAtrue
 \global\setbox\ZoneABOX=\box\VOIDBOX
 \global\setbox\ZoneBBOX=\box\VOIDBOX
 \global\setbox\ZoneCBOX=\box\VOIDBOX
 \loop
  \ifMakePageInComplete
 \FindNext
 \ifnum\StackPointer=\m@ne
  \NextItem=\m@ne
  \MoreFiguresfalse
  \MoreTablesfalse
 \fi
 \ifnum\NextItem=\Figure
   \FindItem{\Figure}{\NextFigure}
   \ifnum\StackPointer=\m@ne \global\MoreFiguresfalse
   \else
    \GetItemSPAN{\StackPointer}
    \ifnum\ItemSPAN=\Single \def\Zone{\InZoneB}\relax
     \ifFigInZoneC \global\MoreFiguresfalse\fi
    \else
     \def\Zone{\InZoneA}
     \ifFigInZoneB \def\Zone{\InZoneC}\fi
    \fi
   \fi
   \ifMoreFigures\Print{}\FigureItems\fi
 \fi
\ifnum\NextItem=\Table
   \FindItem{\Table}{\NextTable}
   \ifnum\StackPointer=\m@ne \global\MoreTablesfalse
   \else
    \GetItemSPAN{\StackPointer}
    \ifnum\ItemSPAN=\Single\relax
     \ifTabInZoneC \global\MoreTablesfalse\fi
    \else
     \def\Zone{\InZoneA}
     \ifTabInZoneB \def\Zone{\InZoneC}\fi
    \fi
   \fi
   \ifMoreTables\Print{}\TableItems\fi
 \fi
   \MakePageInCompletefalse 
   \ifMoreFigures\MakePageInCompletetrue\fi
   \ifMoreTables\MakePageInCompletetrue\fi
 \repeat
 \ifZoneAFullPage
  \global\TextSize=\z@
  \global\ZoneBSize=\z@
  \global\vsize=\z@\relax
  \global\topskip=\z@\relax
  \vbox to \z@{\vss}
  \eject
 \else
 \global\advance\ZoneBSize by -\ZoneBAdjust
 \global\vsize=\ZoneBSize
 \global\hsize=\ColumnWidth
 \global\ZoneBAdjust=\z@
 \ifdim\TextSize<23pt
 \Warn{}
 \Warn{* Making column fall short: TextSize=\the\TextSize *}
 \vskip-\lastskip\eject\fi
 \fi
}

\def\MakeRightCol{
 \global\TextSize=\ZoneBSize
 \MakePageInCompletetrue
 \MoreFigurestrue
 \MoreTablestrue
 \global\FirstSingleItemtrue
 \global\setbox\ZoneBBOX=\box\VOIDBOX
 \def\Zone{\InZoneB}
 \loop
  \ifMakePageInComplete
 \FindNext
 \ifnum\StackPointer=\m@ne
  \NextItem=\m@ne
  \MoreFiguresfalse
  \MoreTablesfalse
 \fi
 \ifnum\NextItem=\Figure
   \FindItem{\Figure}{\NextFigure}
   \ifnum\StackPointer=\m@ne \MoreFiguresfalse
   \else
    \GetItemSPAN{\StackPointer}
    \ifnum\ItemSPAN=\Double\relax
     \MoreFiguresfalse\fi
   \fi
   \ifMoreFigures\Print{}\FigureItems\fi
 \fi
 \ifnum\NextItem=\Table
   \FindItem{\Table}{\NextTable}
   \ifnum\StackPointer=\m@ne \MoreTablesfalse
   \else
    \GetItemSPAN{\StackPointer}
    \ifnum\ItemSPAN=\Double\relax
     \MoreTablesfalse\fi
   \fi
   \ifMoreTables\Print{}\TableItems\fi
 \fi
   \MakePageInCompletefalse 
   \ifMoreFigures\MakePageInCompletetrue\fi
   \ifMoreTables\MakePageInCompletetrue\fi
 \repeat
 \ifZoneAFullPage
  \global\TextSize=\z@
  \global\ZoneBSize=\z@
  \global\vsize=\z@\relax
  \global\topskip=\z@\relax
  \vbox to \z@{\vss}
  \eject
 \else
 \global\vsize=\ZoneBSize
 \global\hsize=\ColumnWidth
 \ifdim\TextSize<23pt
 \Warn{}
 \Warn{* Making column fall short: TextSize=\the\TextSize *}
 \vskip-\lastskip\eject\fi
\fi
}

\def\FigureItems{
 \Print{Considering...}
 \ShowItem{\StackPointer}
 \GetItemBOX{\StackPointer} 
 \GetItemSPAN{\StackPointer}
  \CheckFitInZone 
  \ifnum\ItemFits=\Yes
   \ifnum\ItemSPAN=\Single
     \ChangeStatus{\StackPointer}{\InZoneB} 
     \global\FigInZoneBtrue
     \ifFirstSingleItem
      \hbox{}\vskip-\BodgeHeight
     \global\advance\ItemSIZE by \TextLeading
     \fi
     \unvbox\ItemBOX\ItemSep
     \global\FirstSingleItemfalse
     \global\advance\TextSize by -\ItemSIZE
     \global\advance\TextSize by -\TextLeading
   \else
    \ifFirstZoneA
     \global\advance\ItemSIZE by \TextLeading
     \global\FirstZoneAfalse\fi
    \global\advance\TextSize by -\ItemSIZE
    \global\advance\TextSize by -\TextLeading
    \global\advance\ZoneBSize by -\ItemSIZE
    \global\advance\ZoneBSize by -\TextLeading
    \ifFigInZoneB\relax
     \else
     \ifdim\TextSize<3\TextLeading
     \global\ZoneAFullPagetrue
     \fi
    \fi
    \ChangeStatus{\StackPointer}{\Zone}
    \ifnum\Zone=\InZoneC \global\FigInZoneCtrue\fi
  \fi
   \Print{TextSize=\the\TextSize}
   \Print{ZoneBSize=\the\ZoneBSize}
  \global\advance\NextFigure \@ne
   \Print{This figure has been placed.}
  \else
   \Print{No space available for this figure...holding over.}
   \Print{}
   \global\MoreFiguresfalse
  \fi
}

\def\TableItems{
 \Print{Considering...}
 \ShowItem{\StackPointer}
 \GetItemBOX{\StackPointer} 
 \GetItemSPAN{\StackPointer}
  \CheckFitInZone 
  \ifnum\ItemFits=\Yes
   \ifnum\ItemSPAN=\Single
    \ChangeStatus{\StackPointer}{\InZoneB}
     \global\TabInZoneBtrue
     \ifFirstSingleItem
      \hbox{}\vskip-\BodgeHeight
     \global\advance\ItemSIZE by \TextLeading
     \fi
     \unvbox\ItemBOX\ItemSep
     \global\FirstSingleItemfalse
     \global\advance\TextSize by -\ItemSIZE
     \global\advance\TextSize by -\TextLeading
   \else
    \ifFirstZoneA
    \global\advance\ItemSIZE by \TextLeading
    \global\FirstZoneAfalse\fi
    \global\advance\TextSize by -\ItemSIZE
    \global\advance\TextSize by -\TextLeading
    \global\advance\ZoneBSize by -\ItemSIZE
    \global\advance\ZoneBSize by -\TextLeading
    \ifFigInZoneB\relax
     \else
     \ifdim\TextSize<3\TextLeading
     \global\ZoneAFullPagetrue
     \fi
    \fi
    \ChangeStatus{\StackPointer}{\Zone}
    \ifnum\Zone=\InZoneC \global\TabInZoneCtrue\fi
   \fi
  \global\advance\NextTable \@ne
   \Print{This table has been placed.}
  \else
  \Print{No space available for this table...holding over.}
   \Print{}
   \global\MoreTablesfalse
  \fi
}


\def\CheckFitInZone{%
{\advance\TextSize by -\ItemSIZE
 \advance\TextSize by -\TextLeading
 \ifFirstSingleItem
  \advance\TextSize by \TextLeading
 \fi
 \ifnum\Zone=\InZoneA\relax
  \else \advance\TextSize by -\ZoneBAdjust
 \fi
 \ifdim\TextSize<3\TextLeading \global\ItemFits=\No
 \else \global\ItemFits=\Yes\fi}
}

\def\BeginOpening{%
  \ninepoint
  \thispagestyle{titlepage}%
  \global\setbox\ItemBOX=\vbox\bgroup%
    \hsize=\PageWidth%
    \hrule height \z@
    \ifsinglecol\vskip 6pt\fi 
}

\let\begintopmatter=\BeginOpening  

\def\EndOpening{%
  \One
  \egroup
  \ifsinglecol
    \box\ItemBOX%
    \vskip\TextLeading plus 2\TextLeading
    \@noafterindent
  \else
    \ItemNUMBER=\z@%
    \ItemTYPE=\Figure
    \ItemSPAN=\Double
    \ItemSTATUS=\InStack
    \JoinStack
  \fi
}


\newif\if@here  \@herefalse

\def\no@float{\global\@heretrue}
\let\nofloat=\relax 

\def\beginfigure{%
  \@ifstar{\global\@dfloattrue \@bfigure}{\global\@dfloatfalse \@bfigure}%
}

\def\@bfigure#1{%
  \par
  \if@dfloat
    \ItemSPAN=\Double
    \TEMPDIMEN=\PageWidth
  \else
    \ItemSPAN=\Single
    \TEMPDIMEN=\ColumnWidth
  \fi
  \ifsinglecol
    \TEMPDIMEN=\PageWidth
  \else
    \ItemSTATUS=\InStack
    \ItemNUMBER=#1%
    \ItemTYPE=\Figure
  \fi
  \bgroup
    \hsize=\TEMPDIMEN
    \global\setbox\ItemBOX=\vbox\bgroup
      \eightpoint\nostb@ls{10pt}%
      \let\caption=\fig@caption
      \ifsinglecol \let\nofloat=\no@float\fi
}

\def\fig@caption#1{%
  \vskip 5.5pt plus 6pt%
  \bgroup 
    \eightpoint\nostb@ls{10pt}%
    \setbox\TEMPBOX=\hbox{#1}%
    \ifdim\wd\TEMPBOX>\TEMPDIMEN
      \noindent \unhbox\TEMPBOX\par
    \else
      \hbox to \hsize{\hfil\unhbox\TEMPBOX\hfil}%
    \fi
  \egroup
}

\def\endfigure{%
  \par\egroup 
  \egroup
  \ifsinglecol
    \if@here \midinsert\global\@herefalse\else \topinsert\fi
      \unvbox\ItemBOX
    \endinsert
  \else
    \JoinStack
    \Print{Processing source for figure \the\ItemNUMBER}%
  \fi
}


\newbox\tab@cap@box
\def\tab@caption#1{\global\setbox\tab@cap@box=\hbox{#1\par}}

\newtoks\tab@txt@toks
\long\def\tab@txt#1{\global\tab@txt@toks={#1}\global\table@txttrue}

\newif\iftable@txt  \table@txtfalse
\newif\if@dfloat    \@dfloatfalse

\def\begintable{%
  \@ifstar{\global\@dfloattrue \@btable}{\global\@dfloatfalse \@btable}%
}

\def\@btable#1{%
  \par
  \if@dfloat
    \ItemSPAN=\Double
    \TEMPDIMEN=\PageWidth
  \else
    \ItemSPAN=\Single
    \TEMPDIMEN=\ColumnWidth
  \fi
  \ifsinglecol
    \TEMPDIMEN=\PageWidth
  \else
    \ItemSTATUS=\InStack
    \ItemNUMBER=#1%
    \ItemTYPE=\Table
  \fi
  \bgroup
    \eightpoint\nostb@ls{10pt}%
    \global\setbox\ItemBOX=\vbox\bgroup
      \let\caption=\tab@caption
      \let\tabletext=\tab@txt
      \ifsinglecol \let\nofloat=\no@float\fi
}

\def\endtable{%
  \par\egroup 
  \egroup
  \setbox\TEMPBOX=\hbox to \TEMPDIMEN{%
    \eightpoint\nostb@ls{10pt}%
    \hss
    \vbox{%
      \hsize=\wd\ItemBOX
      \ifvoid\tab@cap@box
      \else
        \noindent\unhbox\tab@cap@box
        \vskip 5.5pt plus 6pt%
      \fi
      \box\ItemBOX
      \iftable@txt
        \vskip 10pt%
        \noindent\the\tab@txt@toks
        \global\table@txtfalse
      \fi
    }%
    \hss
  }%
  \ifsinglecol
    \if@here \midinsert\global\@herefalse\else \topinsert\fi
      \box\TEMPBOX
    \endinsert
  \else
    \global\setbox\ItemBOX=\box\TEMPBOX
    \JoinStack
    \Print{Processing source for table \the\ItemNUMBER}%
  \fi
}

\def\UnloadZoneA{%
\FirstZoneAtrue
 \Iteration=\z@
  \loop
   \ifnum\Iteration<\LengthOfStack
    \GetItemSTATUS{\Iteration}
    \ifnum\ItemSTATUS=\InZoneA
     \GetItemBOX{\Iteration}
     \ifFirstZoneA \vbox to \BodgeHeight{\vfil}%
     \FirstZoneAfalse\fi
     \unvbox\ItemBOX\ItemSep
     \LeaveStack{\Iteration}
     \else
     \advance\Iteration \@ne
   \fi
 \repeat
}

\def\UnloadZoneC{%
\Iteration=\z@
  \loop
   \ifnum\Iteration<\LengthOfStack
    \GetItemSTATUS{\Iteration}
    \ifnum\ItemSTATUS=\InZoneC
     \GetItemBOX{\Iteration}
     \ItemSep\unvbox\ItemBOX
     \LeaveStack{\Iteration}
     \else
     \advance\Iteration \@ne
   \fi
 \repeat
}


\def\ShowItem#1{
  {\GetItemAll{#1}
  \Print{\the#1:
  {TYPE=\ifnum\ItemTYPE=\Figure Figure\else Table\fi}
  {NUMBER=\the\ItemNUMBER}
  {SPAN=\ifnum\ItemSPAN=\Single Single\else Double\fi}
  {SIZE=\the\ItemSIZE}}}
}

\def\ShowStack{%
 \Print{}
 \Print{LengthOfStack = \the\LengthOfStack}
 \ifnum\LengthOfStack=\z@ \Print{Stack is empty}\fi
 \Iteration=\z@
 \loop
 \ifnum\Iteration<\LengthOfStack
  \ShowItem{\Iteration}
  \advance\Iteration \@ne
 \repeat
}

\def\B#1#2{%
\hbox{\vrule\kern-0.4pt\vbox to #2{%
\hrule width #1\vfill\hrule}\kern-0.4pt\vrule}
}


\newif\ifsinglecol   \singlecolfalse

\def\onecolumn{%
  \global\output={\singlecoloutput}%
  \global\hsize=\PageWidth
  \global\vsize=\PageHeight
  \global\ColumnWidth=\hsize
  \global\TextLeading=12pt
  \global\Leading=12
  \global\singlecoltrue
  \global\let\onecolumn=\relax
  \global\let\footnote=\sing@footnote
  \global\let\vfootnote=\sing@vfootnote
  \ninepoint 
  \message{(Single column)}%
}

\def\singlecoloutput{%
  \shipout\vbox{\PageHead\vbox to \PageHeight{\pagebody\vss}\PageFoot}%
  \advancepageno
  \ifplate@page
    \shipout\vbox{%
      \sp@pagetrue
      \def\sp@type{plate}%
      \global\plate@pagefalse
      \PageHead\vbox to \PageHeight{\unvbox\plt@box\vfil}\PageFoot%
    }%
    \message{[plate]}%
    \advancepageno
  \fi
  \ifnum\outputpenalty>-\@MM \else\dosupereject\fi%
}

\def\ItemSep{\vskip\ItemSepamount\relax}

\def\ItemSepbreak{\par\ifdim\lastskip<\ItemSepamount
  \removelastskip\penalty-200\ItemSep\fi%
}


\let\@@endinsert=\endinsert 

\def\endinsert{\egroup 
  \if@mid \dimen@\ht\z@ \advance\dimen@\dp\z@ \advance\dimen@12\p@
    \advance\dimen@\pagetotal \advance\dimen@-\pageshrink
    \ifdim\dimen@>\pagegoal\@midfalse\p@gefalse\fi\fi
  \if@mid \ItemSep\box\z@\ItemSepbreak
  \else\insert\topins{\penalty100 
    \splittopskip\z@skip
    \splitmaxdepth\maxdimen \floatingpenalty\z@
    \ifp@ge \dimen@\dp\z@
    \vbox to\vsize{\unvbox\z@\kern-\dimen@}
    \else \box\z@\nobreak\ItemSep\fi}\fi\endgroup%
}


\def\gobbleone#1{}
\def\gobbletwo#1#2{}
\let\footnote=\gobbletwo 
\let\vfootnote=\gobbleone

\def\sing@footnote#1{\let\@sf\empty 
  \ifhmode\edef\@sf{\spacefactor\the\spacefactor}\/\fi
  \hbox{$^{\hbox{\eightpoint #1}}$}\@sf\sing@vfootnote{#1}%
}

\def\sing@vfootnote#1{\insert\footins\bgroup\eightpoint\b@ls{9pt}%
  \interlinepenalty\interfootnotelinepenalty
  \splittopskip\ht\strutbox 
  \splitmaxdepth\dp\strutbox \floatingpenalty\@MM
  \leftskip\z@skip \rightskip\z@skip \spaceskip\z@skip \xspaceskip\z@skip
  \noindent $^{\scriptstyle\hbox{#1}}$\hskip 4pt%
    \footstrut\futurelet\next\fo@t%
}

\def\footnoterule{\kern-3\p@ \hrule height \z@ \kern 3\p@}

\skip\footins=19.5pt plus 12pt minus 1pt
\count\footins=1000
\dimen\footins=\maxdimen

\def\note#1#2{%
  \let\@sf=\empty \ifhmode\edef\@sf{\spacefactor\the\spacefactor}\/\fi
  #1\insert\footins\bgroup
    \eightpoint\b@ls{10pt}\rm
    \interlinepenalty\interfootnotelinepenalty
    \splitmaxdepth\dp\strutbox \floatingpenalty\@MM
    \leftskip\z@skip \rightskip\z@skip \spaceskip\z@skip \xspaceskip\z@skip
    \noindent\footstrut #1$\,$#2\strut\par
  \egroup
  \@sf\relax}


\def\landscape{%
  \global\TEMPDIMEN=\PageWidth
  \global\PageWidth=\PageHeight
  \global\PageHeight=\TEMPDIMEN
  \global\let\landscape=\relax
  \onecolumn
  \message{(landscape)}%
  \raggedbottom
}


\output{%
  \ifLeftCOL
    \global\setbox\LeftBOX=\vbox to \ZoneBSize{\box255\unvbox\ZoneBBOX
      \ifvoid\footins\else
        \vskip\skip\footins\unvbox\footins\fi
    }%
    \global\LeftCOLfalse
    \MakeRightCol
  \else
    \setbox\RightBOX=\vbox to \ZoneBSize{\box255\unvbox\ZoneBBOX
      \ifvoid\footins\else
        \vskip\skip\footins\unvbox\footins\fi
    }%
    \setbox\MidBOX=\hbox{\box\LeftBOX\hskip\ColumnGap\box\RightBOX}%
    \setbox\PageBOX=\vbox to \PageHeight{%
      \UnloadZoneA\box\MidBOX\UnloadZoneC}%
    \shipout\vbox{\PageHead\vbox to \PageHeight{\box\PageBOX\vss}\PageFoot}%
    \advancepageno
    \ifplate@page
      \shipout\vbox{%
        \sp@pagetrue
        \def\sp@type{plate}%
        \global\plate@pagefalse
        \PageHead\vbox to \PageHeight{\unvbox\plt@box\vfil}\PageFoot%
      }%
      \message{[plate]}%
      \advancepageno
    \fi
    \global\LeftCOLtrue
    \CleanStack
    \MakePage
  \fi
}


\Warn{\start@mess}

\newif\ifCUPmtplainloaded 
\ifprod@font
  \global\CUPmtplainloadedtrue
\fi

\def\mnmacrosloaded{} 

\catcode `\@=12 



\fi

\newif\ifAMStwofonts

\ifCUPmtplainloaded \else
  \NewTextAlphabet{textbfit} {cmbxti10} {}
  \NewTextAlphabet{textbfss} {cmssbx10} {}
  \NewMathAlphabet{mathbfit} {cmbxti10} {} 
  \NewMathAlphabet{mathbfss} {cmssbx10} {} 
  \ifAMStwofonts
    \NewSymbolFont{upmath} {eurm10}
    \NewSymbolFont{AMSa} {msam10}
    \NewMathSymbol{\upi}     {0}{upmath}{19}
    \NewMathSymbol{\umu}     {0}{upmath}{16}
    \NewMathSymbol{\upartial}{0}{upmath}{40}
    \NewMathSymbol{\leqslant}{3}{AMSa}{36}
    \NewMathSymbol{\geqslant}{3}{AMSa}{3E}

    \let\leq=\leqslant 
    \let\geq=\geqslant 
  \else
    \def\umu{\mu}
    \def\upi{\pi}
    \def\upartial{\partial}
  \fi
\fi


\pageoffset{-2.5pc}{0pc}

\loadboldmathnames



\def\refrule{\hbox to 3pc{\leaders\hrule depth-2pt height 2.4pt\hfill}}
\def\ref{\hangindent=1truecm}

\def\araa{ARA\&A\,}

\def\aj{AJ\,}

\def\aea{A\&A\,}

\def\apj{ApJ\,}

\def\apjs{ApJS\,}

\def\mnras{MNRAS\,}

\def\etal{et al.$\,\,$}
\def\lb{L_{\rm B}}
\def\lx{L_{\rm X}}

\def\vr{v_{\rm rot}}
\def\sgc{\sigma _{\rm c}}
\def\Re{R_{\rm e}}
\def\aef{a_{\rm e}}

\def\kms{{\rm\,km\,s^{-1}}}

\def\lsun{{L$_\odot$}}
\def\as{$^{\prime\prime}$\hskip-4pt .\hskip0.5pt}
\def\am{$^{\prime}$\hskip-3pt .}
\catcode`\@=11
\def\gsim{\ifmmode{\mathrel{\mathpalette\@versim>}}
    \else{$\mathrel{\mathpalette\@versim>}$}\fi}
\def\lsim{\ifmmode{\mathrel{\mathpalette\@versim<}}
    \else{$\mathrel{\mathpalette\@versim<}$}\fi}
\def\@versim#1#2{\lower 2.9truept \vbox{\baselineskip 0pt \lineskip
    0.5truept \ialign{$\m@th#1\hfil##\hfil$\crcr#2\crcr\sim\crcr}}}
\catcode`\@=12

\begintopmatter 

\title{X-ray emission and internal kinematics in early-type galaxies.
I. Observations$^{\star}$}
\author{S.~Pellegrini,$^1$ E.~V.~Held,$^{2,3}$ and L.~Ciotti$^2$}
\affiliation{$^1$ Dipartimento di Astronomia, Universit\`a di Bologna,
via Zamboni 33, I-40126 Bologna, Italy}
\smallskip
\affiliation{$^2$ Osservatorio Astronomico di Bologna, 
via Zamboni 33, I-40126 Bologna, Italy}
\smallskip
\affiliation{$^3$ Osservatorio Astronomico di Padova, vicolo dell'Osservatorio 
5, I-35122 Padova, Italy}

\shortauthor{S. Pellegrini et al.}
\shorttitle{X-ray emission and internal kinematics}
        

\abstract {Long slit spectroscopic data for 7 early-type galaxies
with X-ray emission have been analyzed to derive velocity dispersion and 
radial velocity profiles. Major axis rotation curves out to $R\sim\Re$ are
presented. Adding these new data to those available in the literature, we have
built a sample of early-type galaxies with detected X-ray emission 
and known kinematics (central velocity dispersion $\sgc $ and maximum 
rotational velocity $\vr$). This sample is used to investigate from an 
observational point of view
the role of rotation and flattening on the X-ray emission, particularly with 
regard to the X-ray underluminosity of flat systems. 

The trend between the X-ray to optical ratio $\lx/\lb$, a measure of the hot 
gas content of the galaxies, and $\vr/\sgc$ is L-shaped, with the 
X-ray brightest objects confined at $\vr/\sgc\lsim 0.4$, both for Es and S0s.
Neither for low or intermediate, nor for high $\lx/\lb$, 
there is any clear correlation between X-ray emission and rotational 
properties. The trend between $\lx/\lb$ and the ellipticity $\epsilon$ is also 
L-shaped, and resembles that between $\lx/\lb$ and $\vr/\sgc$:
there are no high $\lx/\lb$ objects with high $\epsilon$. The relationships 
between $\lx/\lb$, the anisotropy parameter $(v/\sigma)^*$, and the isophotal 
shape parameter $a_4/a$ have also been investigated, but no significant trends 
have been found. 

The existence of a relation between $\vr/\sgc$ and $\epsilon$ makes it 
difficult to assess on a purely observational ground whether rotation or 
flattening is at the basis of the L-shaped trends found, 
although the trend with 
$\vr/\sgc$ is sharper than that with $\epsilon$.  Our observational findings
are then discussed in connection with the effects that rotation and 
flattening are predicted to have in the steady state cooling flow and in the 
evolutionary wind/outflow/inflow scenarios for the hot gas behavior.
 }

\keywords {galaxies: elliptical and lenticular, cD -- galaxies: ISM --
galaxies: kinematics and dynamics -- galaxies: structure -- X-rays: galaxies}

\maketitle 

\section{Introduction}

\tx X-ray observations, beginning with the $Einstein $ Observatory, have 
demonstrated that normal early-type galaxies are X-ray emitters, with  0.2--4 
keV luminosities ranging from $\sim 10^{40}$ to $\sim 10^{43}$ erg s$^{-1}$ 
(Fabbiano 1989; Fabbiano, Kim \& Trinchieri 1992). The X-ray luminosity $\lx$ 
is found to correlate with the blue luminosity $\lb$ ($\lx\propto 
\lb^{2.0\pm 0.2}$), although there is a large scatter of roughly two 
orders of
magnitude in $\lx$ at any fixed $\lb >3\times 10^{10}$\lsun. The observed X-ray
spectra of high $\lx/\lb$ galaxies are consistent with thermal emission from 
hot, optically thin gas, while those of low $\lx/\lb$ objects can be mostly 
accounted for by emission from stellar sources (Kim, Fabbiano \& Trinchieri 
1992).\note {}{$^{\star}$Based on observations collected at the
European Southern Observatory, La Silla, Chile} 
The scatter in the $\lx -\lb$ diagram has been recognized as the most 
striking feature of the X-ray properties of early-type galaxies. It was
originally explained in terms of environmental differences (e.g., 
White \& Sarazin 1991), or in terms of different 
dynamical phases for the hot gas flows, ranging from winds to subsonic outflows
to inflows, going from the lowest to the highest $\lx/\lb$ values (Ciotti 
\etal 1991, hereafter WOI scenario).

Recent observational results, followed by new theoretical work, have produced a
new debate on the explanation of the scatter. Eskridge, Fabbiano \& Kim 
(1995a,b; hereafter EFKa,b) conducted a multivariate statistical analysis of data measuring the 
optical and X-ray properties of the {\it Einstein} sample of early-type 
galaxies, the largest X-ray-selected sample of such galaxies presently 
available. They showed that on average S0 galaxies have lower $\lx$ and 
$\lx/\lb$ at any fixed $\lb$ than do ellipticals. Moreover they found that 
galaxies with axial ratio close to unity span the full range of $\lx$, while 
flat systems all have $\lx\lsim 10^{41}$ erg s$^{-1}$. The relationship defined
by $\lx/\lb$ is stronger than that defined by $\lx$, and it is in the sense 
that at any fixed $\lb$ the X-ray-brightest galaxies are also the roundest. 
This correlation holds for both morphological subsets of Es and S0s. 

These new observational findings have been followed by theoretical studies 
concerning the role played by the shape of the mass distribution and by 
galactic rotation in producing a different $\lx$, at fixed $\lb$ (Kley \& 
Mathews 1995; Ciotti \& Pellegrini 1996, hereafter CP; Brighenti \& Mathews 
1996, hereafter BM; D'Ercole \& Ciotti 1996). The different views of 
these authors, and their different predictions, will be presented in Section 6.

In this paper we focus on the effect of galactic rotation on $\lx$, from an 
observational point of view. We have built the first sample of early-type 
galaxies for which $\lx$, the maximum rotational velocity $\vr$, and the 
central velocity dispersion $\sgc$ are known, collecting data from the
literature, and performing ourselves new long-slit spectroscopic observations 
for 7 galaxies. The resulting sample consists of 52 galaxies. In a subsequent 
paper (Paper II) we build the mass models specific for two galaxies with 
kinematic and photometric data obtained by us, and then we perform
hydrodynamical simulations for such models.

This paper is organized as follows. In Section 2 we present our sample, in 
Section 3 we describe the details of the observations, in Section 4 we briefly
summarize the reduction and analysis procedures, in Section 5 we present the
results of our observations, in Section 6 we discuss the different theoretical 
scenarios in relation with the observational data, and in Section 7 we present
our conclusions.

\begintable*{1}
\caption{{\bf Table 1.} Observed Galaxies }
\halign{
\rm#\hfil& #\hfil &\hfil#\hfil & \hfil#\hfil & \hfil#\hfil & \hfil#
\hfil & \hfil#\hfil & \hfil#\hfil & \hfil#\hfil & \hfil#\hfil &\hfil#\hfil &
\quad\quad\hfil#\hfil \cr
Galaxy &$\quad\quad$ Type &$\,\,$ B & $\,$ log($\lb$)$\,$ & $\,$ log($\lx$) 
$\,$ & $\,$ log($\lx/\lb$)$\,$ & $\sgc$ &$\,$ PA $\,$ & $\,\,\,\,$
 d $\,\,\,\,\,$  & $\,$ $\Re$ &\quad $b/a$        \cr
   &$\quad$ RSA,$\,\,$RC3 &$\,\,$ mag& $L_{\odot}$& erg s$^{-1}$ &erg s$^
{-1}/L_{\odot}$ & km s$^{-1}$ & deg &$\,$ Mpc
$\,\,\,\,$ & $\,\,\,\,$ $^{\prime\prime}$ & \cr
$\quad$(1) & $\quad\quad\quad$(2) &$\,\,$ (3) & (4) & (5) & (6) & (7) & (8)  &
 (9)  & (10)  &\quad(11) \cr
\noalign{\vskip 10 pt}
 NGC~2563 &$\quad\quad$   EL$^a$, S0&$\,\,$ 13.01 & 10.95 & 42.12 & 31.17 & 260 
& 80    & 96.1 & 19.37 & \quad 0.72 \cr
 NGC~3078 &$\quad\quad$   E3, E2+   &$\,\,$ 11.86 & 10.91 & 41.17 & 30.26 & 237
& 177   & 53.7 & 22.76 & \quad 0.83 \cr
 NGC~3258 &$\quad\quad$   E1, E1    &$\,\,$ 12.14 & 10.90 & 41.48 & 30.58 & 271 
&  75   & 60.3 & 30.00 & \quad 0.85 \cr
 NGC~3923 & $\,\,$ E4/S0, E4--5     &$\,\,$ 10.27 & 11.32 & 41.35 & 30.03 & 241 
&  50   & 41.5 & 49.79 & \quad 0.66 \cr
 NGC~4526 &$\quad\quad$   S0, S0    &$\,\,$ 10.54 & 10.84 & 40.14 & 29.30 & 267 
& 113   & 27.0 & 44.37 & \quad 0.33 \cr
 NGC~4753 &$\quad\quad$   S0, I0    &$\,\,$ 10.53 & 10.75 & 40.08 & 29.33 &    
&  80   & 24.3 & 45.41 & \quad 0.47 \cr
 NGC~4756 &$\quad\quad$   E3, SAB0  &$\,\,$ 13.27 & 10.78 & 42.06 & 31.28 &    
&  50   & 88.4 & 16.49 & \quad 0.49 \cr
}
\tabletext{Notes -- Col. (2) gives the classifications of the galaxies from 
Sandage \& Tammann (1987) (RSA), and from de Vaucouleurs \etal (1991) (RC3); 
the total B magnitude in col. (3) is $B_{\rm T}^0$ from RC3. $\lb$ in col. (4)
has been derived using distances listed in col. (9), which are given in 
Fabbiano \etal (1992); these are taken from Tully (1988), or are estimated from
the CfA redshift survey data; a Hubble constant of 50 km s$^{-1}$ Mpc$^{-1}$ is
assumed. The X-ray luminosity in col. (5) is from Fabbiano \etal (1992), and 
$\lx/\lb$ in col. (6) has been estimated from cols. (4) and (5). The central 
velocity dispersion in col. (7) is from the catalog of McElroy (1995). The position angle of 
the major axis in col. (8) is from RC3, as are the effective radius in col. 
(10), and the minor to major axes ratio in col. (11). $^a$ EL = elliptical or 
lenticular.}
\endtable

\beginfigure*{1}
\vskip 3truecm
\caption{{\bf Figure 1.} The radial velocity profiles and velocity dispersion
         profiles, in km s$^{-1}$, for the observed galaxies. The profiles are
         folded with respect to the center. Error bars represent the formal 
         error estimate of the Fourier Fitting program. Thick marks on the 
         upper abscissa axes in the $\sigma$'s plots are placed at 0.5$\aef$,
that is derived from $\Re$ of Table 1
using the formula $\Re^2=(1-\epsilon)\aef^2$, where $\epsilon=1-b/a$, and 
constant ellipticity with radius is assumed. 
         Open symbols are data SW of the nucleus, and filled ones NE, for 
         NGC~2563 (major axis), NGC~3258, NGC~3923, NGC~4756, NGC~4753; open symbols
         are SE, and filled ones NW, for NGC~4526, NGC~3078. All the data are 
         tabulated in Appendix A.}
\endfigure

\beginfigure*{2}
\vskip 3truecm
\caption{{\bf Figure 1} -- {\it continued}}
\endfigure

\begintable{2}
\caption{{\bf Table 2.} Instrumental parameters }
\halign{%
\rm#\hfil& \quad\rm#\hfil &\qquad\hfil\rm# \cr
\noalign{\vskip 10 pt}
 CCD characteristics: & \cr
 CCD ................................    &  \#24 FA 2048L (1pix=15$\mu$) \cr
 Read-out-noise ................         &  5.24 e$^-$ pix$^{-1}$        \cr
 Conversion factor ............          &  2.90 e$^-$/ADU               \cr
\noalign{\vskip 5 pt}
 SLIT characteristics: & \cr
 spatial scale .....................     & 0\as81 pix$^{-1}$ \cr
 slit length ........................    & 4\am1                \cr
 slit width .........................    & 2$^{\prime\prime}$            \cr
 grating ............................    & $\#$10 2nd order              \cr
 dispersion ........................     & 65.1 \AA$\,$ mm$^{-1}$        \cr
 spectral range ..................       & 4300--6300 \AA                \cr
 resolution -- FWHM ........             & 1.8\AA                   \cr
 seeing ..............................   & 1\as1--1\as4                  \cr
 resolution$^a$ $\sigma_i$...................    & $\sim 50$ km s$^{-1}$ \cr
}
\tabletext {$^a$ Instrumental dispersion, determined as an average of the 
sigma's of Gaussian fits to lines in calibration spectra and to sky emission 
lines, and calculated at 5000 \AA.}
\endtable

\begintable{3}
\caption{{\bf Table 3.} Observations}
\halign{%
\rm#\hfil & \qquad\rm#\hfil & \quad\rm\hfil#   & \quad\rm\hfil#    \cr
 Galaxy   & $t_{int}$       &  date$\,\,\quad$ & PA                \cr
          &  m              &                  & deg               \cr
   (1)    &   (2)           &  (3)$\,\,\quad$  &  (4)              \cr
\noalign{\vskip 10pt}
 NGC~2563    & 40              & 2/25/1995        & $\,\,\,\,$-10$^a$ \cr 
 NGC~2563    & 40              & 2/25/1995        & 80                \cr
 NGC~2563    & 40              & 2/25/1995        & 80                \cr 
 NGC~3078    & 35              & 2/26/1995        & -3                \cr 
 NGC~3258    & 40              & 2/25/1995        & 75                \cr 
 NGC~3258    & 40              & 2/25/1995        & 75                \cr
 NGC~3923    & 35              & 2/26/1995        & 50                \cr
 NGC~4526 SE & 35              & 2/25/1995        & $\,\,\,$75$^b$    \cr
 NGC~4526 SE & 35              & 2/25/1995        & $\,\,\,$75$^b$    \cr
 NGC~4526 NW & 32              & 2/25/1995        & $\,\,\,$75$^c$    \cr
 NGC~4526 NW & 30              & 2/25/1995        & $\,\,\,$75$^c$    \cr
 NGC~4753 SW & 28              & 2/26/1995        & $\,\,\,$80$^d$    \cr
 NGC~4753 SW & 24              & 2/26/1995        & $\,\,\,$80$^d$    \cr
 NGC~4753 NE & 28              & 2/26/1995        & $\,\,\,$80$^e$    \cr
 NGC~4753 NE & 24              & 2/26/1995        & $\,\,\,$80$^e$    \cr
 NGC~4756    & 40              & 2/26/1995        & 50                \cr
 NGC~4756    & 40              & 2/26/1995        & 50                \cr
 }
\tabletext{Notes -- Col. (2) gives the integration time in minutes, col. (3) 
gives the date of the observations, col. (4) gives the input position angle 
(in degrees from N to E) at the telescope. When not specified, the
slit was aligned along the photometric major axis, given in col. (7) in Table 
1. $^a$Minor axis. $^b$The slit contains the nucleus and the SE direction of 
the major axis. $^c$The slit contains the nucleus and the NW direction; $^d$the
nucleus and the SW direction; $^e$the nucleus and the NE direction.}
\endtable

\section {Sample selection}

\tx Our aim is to investigate the role of rotation in the X-ray luminosity of
early-type galaxies. The 
sample of early-type galaxies with X-ray emission detected by the 
{\it Einstein} satellite is made of 67 objects (Fabbiano \etal 1992); the 
radial velocity $v$ has already been measured for 45 of these. We obtained 
measurements of velocity dispersion $\sigma$ and radial velocity $v$ along the
major axis for a sample of 7 more Es and S0s belonging to the Fabbiano \etal 
(1992) catalog, and for which there are no published kinematic data, at least 
describing their rotational properties. These are three S0s (NGC~2563, 
NGC~4526, NGC~4753), three ellipticals (NGC~3078, NGC~3258, NGC~3923), 
and one E/S0
(NGC~4756). The main characteristics of the observed galaxies (classifications,
total magnitudes, effective radii, position angles, etc.) are summarized in 
Table 1. Our goal was to obtain major axis spectra with good S/N ratio down to
$\sim 0.5\Re$, within which typical early-type galaxies reach their maximum 
rotation velocities (e.g., Fried \& Illingworth 1994).

\section{Observations}

\tx Our long-slit spectroscopic observations were taken with the Boller \& 
Chivens spectrograph attached at the Cassegrain focus of the ESO 1.52 m
telescope at La Silla, on February 25 and 26, 1995. Details of the instrumental
setup are given in Table 2. The wavelength resolution (FWHM) is 1.8\AA, 
equivalent to velocity dispersions of $\sim 50 \kms$ at 5000 \AA.
 
The observations are listed in Table 3. The majority of them were made
with the slit centered on the galaxy nucleus, and aligned along the photometric
major axis. In one case, NGC~2563, a spectrum was also taken along the minor
axis. To better sample the most extended galaxies, NGC~4526 and NGC~4753, and to
better estimate the sky background -- which is derived from the same exposure 
made for the galaxies, using a region in their outskirts -- two series of 
exposures were made: in one series the slit included the 
nucleus and one end of the major axis, in the other it included the nucleus and
the opposite end (see also Table 3). Right before and after each galaxy 
exposure a He-Ar arc lamp spectrum was obtained for calibration purposes; the 
instrument proved to be very stable. Flat field frames, dark and bias frames 
were taken during the day. Four different template stars  with spectral types 
K0III to K3III were observed each night.

\section {Reduction and Analysis}

\begintable{4}
\caption{{\bf Table 4.} Rotational velocities and velocity dispersions}
\halign{%
\rm#\hfil & \rm#\hfil &\rm\hfil# & \rm\hfil# & \rm\hfil# & \rm\hfil# &\rm\hfil#
  \cr
Galaxy  \quad  &\quad \quad$\sgc$  & $\langle\sigma\rangle$ & \quad $\Delta R$ 
  & $\vr $ & \quad R                 & $v_{fit}$ \cr
               & \quad km s$^{-1}$ & km s$^{-1}$            & $^{\prime\prime}$
  & km s$^{-1}$
              & $^{\prime\prime}$ &km s$^{-1}$ \cr
   (1)  &  \quad\quad (2)        &   (3)           &  (4)  & (5) & (6) & (7) \cr
\noalign{\vskip 10pt}
 N2563    & 278.9$\pm$9.4  & 263.5$\pm$6.7  & \quad 9--28  & 109$\pm$21 & 25  & 119   \cr 
 N2563$^a$& 260.5$\pm$16.6 & 236.2$\pm$10.7 & \quad -11--7 &   0$\pm$8  &     &       \cr
 N3078    & 263.5$\pm$8.4  & 253.8$\pm$5.8  & \quad 15--34 & 76$\pm$18  & 30  & 93    \cr 
 N3258    & 274.5$\pm$7.2  & 242.2$\pm$5.9  & \quad 3--20  & 40$\pm$6   & 20  & 41    \cr 
 N3923    & 248.8$\pm$6.0  & 244.4$\pm$4.6  & \quad -24--23 & 1$\pm$4   &     &       \cr
 N4526    & 209.8$\pm$2.7  & 190.2$\pm$2.3  & \quad 50--135& 246$\pm$6  & & \cr
 N4753    & 166.9$\pm$3.1  & 149.8$\pm$2.9  & \quad 20--70 & 127$\pm$8  & 60  & 153   \cr
 N4756    & 203.9$\pm$12.0 & 203.8$\pm$8.3  & \quad 4--11  & 34$\pm$13  & 11  & 31    \cr
 }
\tabletext{Notes -- Col. (2) gives the central (i.e., averaged within $0.1
\aef$) velocity dispersion; col. (3) gives the mean velocity dispersion along 
the major axis (also the minor axis for NGC~2563), calculated as average within
$0.5 \aef$. All the data points have been given equal weight.
Col. (4) gives the radial interval
over which the radial velocities have been averaged; the error weighted mean 
rotational velocities over that interval are given in col. (5). The radial 
velocity at the outermost point of the model fit -- in col. (6) -- is given in
col. (7), when this fit is acceptable. $^a$ minor axis.}
\endtable

\subsection {Calibration}

\tx The reduction steps were carried out with the IRAF package, and included:
subtraction of bias, flat-field division, wavelength calibration and correction
for the geometrical distortion (along the dispersion direction), sky 
subtraction, correction for bad pixels and removal of cosmic ray events. 
Wavelength calibration was obtained by fitting a fifth-order polynomial to 
$\sim 25$ spectral lines from the He-Ar arc lamp spectra.
The residual systematic deviations from the 
fitted polynomials were typically less than 0.05 \AA, and were uncorrelated 
from line to line. 

The tilt of the spectrum was measured by mapping the 
centroid of the light distribution along the slit, and a polynomial was fitted.
The geometrical corrections to be applied were found to be small (within 1--2 
pixel); residual distortions are within $\pm 0.05$ pixels for the standard
stars, 
and within $\pm 0.3$ pixel for the galaxies. The end product was a 
two-dimensional spectrum, binned logarithmically in the wavelength direction, 
and linearly in the spatial one. Finally, the night sky spectrum was determined
for each galaxy from the outer ends of each spectrum, and subtracted from the 
image.

\subsection {Radial velocities and velocity dispersions}

\tx The spectra have been analyzed using the Fourier Fitting method (Franx 
\etal 1989) to derive velocity dispersion and radial velocity profiles. This
technique adds spectra along the slit until the specified S/N ratio is 
achieved, subtracts the continuum, filters out of the Fourier transformed 
spectrum the low and high wavenumbers, fits the galaxy spectrum to the stellar
spectrum to determine a gaussian broadening 
function. Changing the template star or the spectral range produced changes in
our results well within the formal errors given by the method.

Independent radial velocities have also been found using the restricted Fourier
method (i.e., with the broadening function kept constant), which is similar to
the cross-correlation method. In the outer regions of the galaxies, whenever 
the S/N was too low for a good determination of $\sigma$, this procedure was 
useful to obtain more secure determinations of $v$. Radial velocities have been
determined out to a radius where the scatter in the results -- when varying the
template star or the parameters of the analysis -- was well within the formal 
errors. 

\section {Results}

\tx The final radial velocity and velocity dispersion profiles for our program
galaxies are plotted in Fig. 1. The kinematic data are tabulated in Appendix A.
The zero points of the radial scales are those that give the maximum symmetry 
of the rotation curves on the two sides of the major axis; these centers are 
always close to the row of maximum intensity in the light profile of the galaxies (when
they differ, the shift is at most of 3 pixels). The solid lines shown in some 
of the radial velocity plots in Fig. 1 are the best fit curves of the shape 
(Franx \etal 1989)
$$v=v_0{R\over r_s+|R|}.\eqno (1) $$
From the best model fit the velocity in the outermost observed region was 
determined; this is tabulated in Table 4. In this table we also list $\vr$, 
i.e., the error weighted mean rotational velocities calculated over the region 
where the rotation curve is mostly flat. Typically we could determine the 
radial velocity profile out to $R\sim\Re$.

Finally, we have derived average dispersions from the center to half effective
major axis $\aef$ (the major axis of the ellipse within which one-half of the 
total B-band flux is emitted). Central velocity dispersions $\sgc$ have been
derived 
within $0.1\aef$. These velocity dispersion values are also listed in Table 4.
The central dispersions derived here can be compared with those derived by 
other authors, for those five galaxies for which such a measure has been made 
(see Table 1). The agreement is excellent for NGC~3258 and NGC~3923, and good 
within the errors for NGC~2563 and NGC~3078: the values of $\sgc$
in Table 1 for these two galaxies have been derived from those of Davies 
\etal 1987, who find respectively $\sgc=261\pm 26$ and $\sgc =238\pm 24$.
The lower $\sgc$ we find for NGC~4526 is due to the larger region used by us to 
estimate it.

\beginfigure{3}
\vskip 3cm
\caption{{\bf Figure 2.} Panel a shows $\lx/\lb$ versus $\vr/\sgc$, with $\lx$
                         in erg s$^{-1}$ from Fabbiano \etal (1992); $\lb$ is 
                         in \lsun, and the values have been derived using 
                         $B_{\rm T}^0$ from RC3, and the distances given in 
                         Fabbiano \etal (1992). In panel b the ratio $\lx/\lb$
                         is plotted against the apparent ellipticity $\epsilon=
                         1-b/a$; $b/a$ values are from RC3. Open circles are 
                         Es, and full circles are S0s.}
\endfigure

Only one galaxy (NGC~3923) does not show significant rotation along the major 
axis; for this object
we calculate $\vr$ as the weighted average of all values within a radius of 
$20^{\prime\prime}$. 

\section {Discussion}

\subsection {X-ray emission, galaxy shape and internal kinematics}

\beginfigure{4}
\vskip 3cm
\caption{{\bf Figure 3.} The relation between $\lx/\lb$ and $\lb$. $\lx$ and 
                         $\lb$ are estimated as in Fig. 2; also the symbols are
                         the same.}
\endfigure

\subsubsection{X-ray emission, rotation and flattening}

\tx We collect in Table 5 the kinematic properties of the Fabbiano \etal (1992)
sample of early-type galaxies with X-ray detection and measured rotational
velocity along the major axis. Including the objects studied in the present 
work, 52 galaxies have both $\sgc$ and $\vr$ measured. This is the first large
sample of early-type galaxies with detected X-ray emission and known 
kinematics, and it is used here to investigate from an observational point of 
view the effect of rotation and flattening on $\lx$.

In Fig. 2 we plot $\lx/\lb$, a measure of the hot gas content of the galaxies,
versus $\vr/\sgc$, an indicator of the importance of rotation, and $\lx/\lb$ 
versus the ellipticity $\epsilon=1-b/a$, for Es and S0s separately. We note the
following:

1) Both morphological types populate the whole range of $\lx/\lb$, i.e., no 
morphological segregation in $\lx/\lb$ is present. S0s, though, are preferentially
found at low or intermediate $\lx/\lb$, while Es are more homogeneously 
distributed (in agreement with the findings of EFKa\note {$^{
\dag}$}{The analysis conducted by EFKa,b is based on 146 
objects, including detections and upper limits on X-ray fluxes.}).

2) The trend between $\lx/\lb$ and $\vr/\sgc$ (Fig. 2a) is L-shaped, with the 
X-ray brightest objects confined at $\vr/\sgc\lsim 0.4$, both for Es and S0s.
As expected, Es are found at lower rotations
with respect to S0s.\note{$^{\ddag}$}
{4 S0s and 6 Es with X-ray detection still lack any measure
of rotation along the major axis (in addition to 5 face-on S0s). These Es
missing in Fig. 2 have $\log(\lx/\lb)>30.2$, and $\lb\gsim 
9\times 10^{10}$\lsun, except one at $\lb=6\times 10^{10}$\lsun. So, they would
probably turn out to be slowly rotating objects, in the upper left portion of
Fig. 2a.} Neither for low or intermediate X-ray to optical ratio, say 
$\log(\lx/\lb)\lsim 30.2$, nor for high $\lx/\lb$, there is any clear 
correlation between X-ray emission and rotational properties. 

3) In our sample of X-ray emitting galaxies the trend discovered by EFKb is still present as an L-shape in Fig. 2b, (reminescent of that in 
Fig. 2a): there are no high $\lx/\lb$ objects with high $\epsilon$. Again, 
no clear correlations with $\epsilon$ are present, just a confinement
of Es at lower flattenings with respect to S0s.

\beginfigure{5}
\vskip 3cm
\caption{{\bf Figure 4.} The relation between $\lx/\lb$ and the anisotropy
parameter $(v/\sigma)^*$ (see Section 6.1.2). Es are open circles and S0s 
full circles. $\langle\sigma\rangle$ has been published for 46 galaxies only;
the plot with $\sgc$ replacing $\langle\sigma\rangle$, and so containing
all the 52 galaxies in our sample, is basically the same.}
\endfigure

\beginfigure{6}
\vskip 3cm
\caption{{\bf Figure 5.} The relation between $\vr/\sgc$ and ellipticity for 
                         the Es (open circles) and S0s (full circles) in our 
                         sample.}
\endfigure

\begintable*{5}
\caption{{\bf Table 5.} $\vr$ and $\sgc$ for the X-ray sample}
\halign{%
\rm#\hfil& \quad\rm\hfil#& \quad\rm\hfil#&\qquad \quad\rm\hfil#&\quad\rm\hfil# 
 & \quad\rm\hfil# \cr
 Galaxy & $\sgc $      &  $\vr $      & \quad Galaxy & $\sgc$ & $\vr$  \cr
        & km s$^{-1}$  & km s$^{-1}$  &        & km s$^{-1}$ & km s$^{-1}$\cr
   (1)  &   (2)        &   (3)        &  (4)   &  (5)   & (6) \cr
\noalign{\vskip 10pt}
NGC~315  &   300 & 15 & NGC~4374 &   296  & 10 \cr
NGC~533  &   301 & 20 & NGC~4382 &   192  & 60 \cr 
NGC~720  &   243 & 27 & NGC~4406 &   250  & 8 \cr
NGC~1052 &   222 & 92 & NGC~4458 &   106  & 20 \cr
NGC~1316 &   243 &135 & NGC~4472 &   303  & 73 \cr
NGC~1332 &   332 &225 & NGC~4473 &   193  & 54 \cr
NGC~1395 &   245 &100 & NGC~4526 &   267  &  246  \cr
NGC~1399 &   308 & 27 & NGC~4552 &   269  &  1.3  \cr 
NGC~1404 &   250 & 90 & NGC~4636 &   207  &  1.7  \cr  
NGC~1407 &   272 & 47 & NGC~4638 &   132  &  130 \cr
NGC~1553 &   200 &150 & NGC~4649 &   339  &   87 \cr      
NGC~1600 &   342 &7.1 & NGC~4697 &   181  &   101 \cr
NGC~2300 &   254 &  5 & NGC~4753 &   167  &   127 \cr
NGC~2563 &   260 &109 & NGC~4756 &   204  &   34  \cr 
NGC~2832 &   310 &  8 & NGC~4762 &   147  &  165 \cr
NGC~2974 &   201 &191 & NGC~5077 &   273  &   15 \cr 
NGC~3078 &   237 & 76 & NGC~5084 &   211  &  200 \cr 
NGC~3258 &   271 & 40 & NGC~5363 &   198  &140 \cr
NGC~3585 &   218 & 45 & NGC~5838 &   290  &225 \cr
NGC~3607 &   241 &108 & NGC~5846 &   252  &    2  \cr  
NGC~3608 &   203 & 43 & NGC~5866 &   174  &  132 \cr 
NGC~3923 &   241 &1.0 & NGC~5982 &   256  & 60 \cr                      
NGC~4168 &   186 & 20 & NGC~7619 &   312  & 65   \cr
NGC~4261 &   326 & 3  & NGC~7626 &   273  & 10 \cr
NGC~4291 &   278 & 68 & IC~4296  &   316  & 36 \cr
NGC~4365 &   261 &  5 & IC~1459  &   316  & 28 \cr  
}
\tabletext{Notes -- Cols. (2) and (5) give the central velocity dispersion 
 from McElroy (1995), cols. (3) and (6) give the rotational velocities. 
 References for cols. (3) and (6) are Bender \etal (1992), Bertola \& 
Capaccioli (1978), Bender \etal (1994), Bertola \etal (1991), Carter (1987), 
 Davies \etal (1983), Davies \& Birkinshaw (1988), 
 D'Onofrio \etal (1995), Dressler \& Sandage (1983), Fried \& Illingworth 
 (1994), Fisher \etal (1995),  Franx \etal (1989), 
 Oosterloo \etal (1994), Prugniel \& Simien (1994), Scorza \& Bender (1995),
 Seifert \& Scorza (1996).}
\endtable

Are the trends of $\lx/\lb$ with $\vr/\sgc$ and $\epsilon$ produced by the
underlying correlation $\lx\propto\lb^2$ (even though this has a very large 
scatter)? It could be that fast rotators and flat galaxies are found at low 
$\lx/\lb$ because they are also preferentially at low $\lb$. Fig. 3 shows that 
this is not the case: in our sample there is no relationship between $\lx/\lb$ 
and $\lb$, neither for Es nor for S0s (as already found by EFKa); moreover, 
S0s and Es span the same range in $\lb$.

Fig. 2 displays projected quantities, not corrected for the inclination of the
galaxies along the line of sight. To test the effect of inclination, we 
correct $\vr/\sgc$ of the S0 subsample using an inclination angle estimated 
on the hypothesis of an intrinsic $b/a=0.25$ (Dressler \& Sandage 1983), and 
that S0s are isotropic rotators, which gives the maximum 
correction on the projected $\vr/\sgc$ (see Binney \& Tremaine 1987). 
After this correction the L-shape in Fig. 2a is preserved, with small shifts
towards higher $\vr/\sgc$ for most of the S0s. The analogous correction
for the ellipticals cannot be made, because there are no obvious ways to derive
even approximate inclination angles.

\subsubsection{X-ray emission, orbital anisotropy and inner isophotal shape}

Another useful investigation is the analysis of the relationship between
X-ray emission, orbital anisotropy, and the presence of inner stellar disks,
that is evidenced by deviations from pure elliptical isophotes (the $a_4/a$
parameter introduced by Bender \etal 1989).
Anisotropy and presence of disks are believed to give information on the 
internal structure of a galaxy and its formation history, and so they might 
have a relationship with the X-ray emission. 
Bender (1990) has discussed the relevance of isophote shape studies for the
understanding of the structure of early-type galaxies. When making a 
statistical analysis of a large sample of objects with log$\lb>10.2$,
no relation between $\lb$ and $\epsilon$, or $\lb$ and anisotropy 
is observed, and only a weak correlation between $a_4/a$ and 
$\lb$ or $\epsilon$ is present, but a clear dependence of anisotropy and 
of $\lx$ on $a_4/a$ is found: disky objects 
are low X-ray emitters, and in general flattened by rotation; boxy and
irregular objects (likely merger remnants, or objects that have suffered 
accretion and interaction events, Nieto 1989) show a large range of $\lx$, and
 various degrees of velocity anisotropy. 
The relationship between X-ray emission and $a_4/a$ has been reanalyzed by EFKb using
their larger sample: the bivariate correlation analysis of $\lb,\,\, \lx,\,\, 
\lx/\lb$ with $a_4/a$ confirms the finding that X-ray luminous galaxies 
tend to be boxier than X-ray faint ones; the trend with $\lx/\lb$, in the 
sense that there is less hot gas in disky objects, is the weakest of the three.

Following Davies \etal (1983), we estimate the amount of anisotropy by calculating the 
quantity $(v/\sigma)^*$, that is the ratio between the observed $\vr/\langle
\sigma\rangle$ and $(\vr/\langle\sigma\rangle)_{\rm OI}$,
which is the value for an oblate 
isotropic rotator with the same ellipticity
($\langle\sigma\rangle$ is the average within $0.5 \Re $).
Although radial changes of $\epsilon$ may produce some inaccuracies in the 
estimate of $(v/\sigma)^*$, in objects with $(v/\sigma)^*\lsim 0.7$ velocity 
dispersion anisotropies are likely to be present (Bender 1988).
In Fig. 4 we show the relationship between $\lx/\lb$ and $(v/\sigma)^*$ for
our sample: there is no clear trend, in particular the L-shape present in Fig.
2a has disappeared.
This is not completely surprising: the effect of using $(v/\sigma)^*$ 
instead of $\vr/\sgc$ is that of 
populating the upper right region of the diagram, because some round, slowly 
rotating objects can have the anisotropy parameter close to unity. On the 
other hand, flat objects with high rotation (i.e., lying in the lower right 
region of Fig. 2a)  might require some anisotropy to maintain their 
flattening, and so they are moved towards the lower left region in Fig. 4.
So, the observations suggest that the major effect on $\lx/\lb$ is
not given by anisotropy, but it is given by the relative 
importance of rotation and random motions.

For what is concerning a possible role of the isophotal shape, $a_4/a$ has been
measured for only 28 galaxies of our sample, and for all of them $a_4/a\cdot 
100\lsim 1.4$. This subsample shows only a marginal trend of $\lx/\lb$ with 
$a_4/a$, in the same sense as that found by EFKb, which is in fact mainly 
produced by those few galaxies with $1.2 < a_4/a\cdot 100 < 2.8$.
Since the subsample with measured $a_4/a$ is limited to $\vr/\sgc \lsim 0.5$,
we cannot discuss whether 
the dynamical and/or structural reason of boxiness and diskiness has some
relationship with the origin of the threshold effects produced by rotation and 
ellipticity in Fig. 2. We can only recall that 
 EFKb do not find a threshold effect as that present in Fig. 2, when 
considering the role of $a_4/a$ for their sample.
Only when a considerably larger sample of galaxies with observed $\lx$ and
$\vr$  will be available, and a robust multivariate analysis will be possible, 
some stronger conclusions can be obtained.

\subsection {Comparison with theoretical predictions}

\tx Recently three different theoretical works have studied the problem of
the effect of rotation, and of the shape of the mass distribution, on X-ray gas
flows. The results of these investigations (CP; BM; D'Ercole \& Ciotti 
1996) are summarized below.

\beginfigure{7}
\vskip 3truecm
\caption{{\bf Figure 6.} The trend with $\vr/\sgc$ of the residuals from the 
                         best fit of the $\lx-\lb$ correlation determined by 
                         EFKa, for our sample (Es are open 
                         circles, S0s are full circles).}
\endfigure

1) Within the WOI scenario that explains the scatter in the $\lx-\lb$ diagram in 
terms of different flow phases for the hot gas, CP relate the gas flow phase (and so $\lx$) to the rotational 
and gravitational properties of the host galaxies. They compare the energy 
budget of the gas in spherical and in flattened/rotating galaxies of the 
same $\lb$ (and so of the same SNIa heating). They find that the flattening 
can be effective in unbounding the hot gas; on the contrary, for a given galaxy
structure, rotation cannot change a bound gaseous halo into an unbound one. So,
CP conclude that S0s and non-spherical Es are less able to retain hot gaseous 
haloes than are rounder systems of the same $\lb$; i.e., more likely they are 
in the outflow or wind phase. This result looks quite attractive because it
can explain the X-ray underluminosity of flatter systems within the class of
{\it both} S0s and Es (the findings of EFKa,b), while ellipticals are slower 
rotators than S0s.  An important consequence of the CP explanation is to
predict {\it not correlation} but {\it segregation} of $\lx/\lb$ with
respect to the flattening, a feature visible in Fig. 2b, at least for S0s. 
This segregation is produced by the fact that at each $\lb$ there is a 
critical $\epsilon$ such that all the flatter objects are in outflow/wind, 
rounder ones are in inflow.

2) Two-dimensional numerical simulations of gas flows in non spherical 
early-type galaxies have been carried out by BM. They consider oblate galaxies
of different axial ratios, sustained by different amounts of ordered and 
disordered kinetic energies, and a present SNIa rate of one tenth of the value 
given by Tammann (1982). The resulting flows are always in the inflow 
phase, and form a massive, cold disk. $\lx$ is reduced in rotating models, 
because the angular momentum prevents the cooling gas from falling directly to
the bottom of the potential well, and the gas cools before entering the 
galactic core region. In this scenario $\lx$ marginally depends on the flatness
of the galaxy (in agreement with the CP calculations applied to the inflow 
case), while models with the same $\lb$ have a range in $\lx/\lb$ of over an 
order of magnitude, going from non rotating objects to E2 galaxies with
$\vr/\sgc=0.66$, or to E4 galaxies with $\vr/\sgc=0.78$. BM so suggest that most of
the spread observed in the $\lx -\lb$ plane could be due to rotation, and that
rotation is the underlying cause of the X-ray underluminosity of flat objects
found by EFKb, because it increases on average with
$\epsilon$, as shown in Fig. 5 for our sample. They predict that $\vr/\sgc$ should correlate
with the residuals in the $\lx -\lb$ plot. We show in Fig. 6 the trend of these
residuals with $\vr/\sgc$: a very weak trend with a large spread is perhaps 
present for S0s.

3) D'Ercole \& Ciotti (1996) perform two-dimensional numerical simulations of 
gas flows for S0s, in the WOI scenario. Luminous and dark matter densities, and
their internal dynamics are the same as in CP. Models without SNIa heating give
qualitatively the BM's results. Models with SNIa at 0.3 the Tammann's rate, 
both with and without rotation, accumulate negligible amounts of cold gas on a
central disk, and very soon develop a partial wind that lasts tens of Gyrs; if
spherical, these models would be in inflow. Rotation decreases the X-ray emission 
slightly (of a factor of two or so), because it just favours the wind,
as predicted by CP. The conclusion is that flat models, rotating or not, can be
significantly less X-ray luminous than spherical ones of the same $\lb$, 
because they are in partial wind when the spherical ones are in inflow; 
rotation has a minor effect.
 
\section {Conclusions}

\tx Which predictions of the previous scenarios find support in our 
observational findings? The existence of a relation between $\vr/\sgc$ and 
$\epsilon$ (Fig. 5) makes it difficult to determine whether rotation or 
flattening produces the trends in Fig. 2. We can at
least consider whether in the data there is some trace of the effect that
rotation and flattening are predicted to have in the pure inflow and in the 
WOI scenarios: flattening has no importance in the first (BM), while it is 
determinant in the second (CP; D'Ercole \& Ciotti 1996).

In the assumption that all galaxies host an inflow, rotation alone cannot
explain the whole observed spread of more than two orders of magnitude in
$\lx$: BM predict a spread of roughly one order of magnitude by requiring
$\vr/\sgc$ up to 0.78, a value quite larger than that shown by the ellipticals
in the X-ray sample (Fig. 2a). Moreover, as discussed in Section 6.2, the expected 
correlation of $\lx/\lb$, or of the residuals of the $\lx -\lb$
correlation, with $\vr/\sgc$ is not present (Fig. 2a and 6).

On the other hand, the trend between $\lx/\lb$ and $\epsilon$ could be in 
agreement with the predictions of CP and D'Ercole \& Ciotti (1996), because of
the threshold effect present in Fig. 2b. The trend of $\lx/\lb$ and $\vr/\sgc$
could be produced by the correlation between $\vr/\sgc$ and $\epsilon$.
It remains unexplained why the L-shape is more pronounced with 
respect to $\vr/\sgc$ rather than to $\epsilon$: below log($\lx/\lb)\approx
30.2$ there are more objects with $\vr/\sgc>0.4$ rather than with $\epsilon
>0.5$.

We have also checked whether there is any relationship in our sample between
X-ray emission, orbital anisotropy, and the presence of inner stellar disks.
We have not found any threshold effect, or significant correlations, between
$\lx/\lb$ and anisotropy or isophotal shape.

So, we conclude that a satisfactory explanation of all the features in Fig. 2
is not given by any of the two scenarios above. A partial interpretation
could be as follows: 

$\bullet$ at low $\lx/\lb$, flat galaxies contribute substantially to the total
number of objects, because of the importance of flattening, that makes the
 partial wind phase favoured. 

$\bullet$ at high $\lx/\lb$, where galaxies are in inflow, the scatter in $\lx$ 
could be due, at least partially, to rotation, as suggested by BM.

Finally, we note that Fig. 2  argues against the hypothesis that some 
peculiar properties of the stellar population of S0s with respect to that of Es
(e.g., a higher SNIa rate, or a younger/bluer stellar population) could be 
responsible for lowering the $\lx/\lb$ ratio: at low rotations and at low 
flattenings, where the two morphological types overlap, S0s span the same range
of $\lx/\lb$ as Es.

\section*{Acknowledgments}

We thank J. Binney, F. Brighenti, A. D'Ercole and G. Galletta for useful 
discussions, and the referee for suggestions that helped improve the paper.

\appendix
\section {kinematic data}

\tx The kinematic results for our program galaxies are tabulated here.
The derived radial 
velocities $v$ and velocity dispersions $\sgc$ along with their formal errors 
$(\pm1 \sigma)$ are given, in km s$^{-1}$. Radii are in arcseconds. The 
position angles in degrees, PA, are consistent with the normal convention, 
increasing from north to east.

\begintable{6}
\caption{NGC~2563 PA 80 major axis }
\halign{%
\rm#\hfil& \qquad\rm#\hfil& \quad\rm\hfil#& \quad\rm\hfil# &\quad\rm\hfil# \cr
 R($^{\prime\prime}$) &  $v$ &  $\Delta v$ & $\sigma$    &$\Delta \sigma $\cr
\noalign{\vskip 10pt}
 -14.3 &  119.6 &  65.9 & 200.0 &  57.7 \cr
  -8.1 &   54.5 &  33.9 & 200.0 &  29.7 \cr
  -5.1 &   98.3 &  39.5 & 248.0 &  50.5 \cr
  -3.0 &  104.1 &  19.7 & 278.4 &  25.2 \cr
  -1.5 &   23.0 &  22.4 & 248.0 &  28.7 \cr
  -0.7 &   -5.3 &  19.1 & 301.7 &  24.6 \cr
   0.1 &  -17.5 &  17.2 & 262.6 &  22.0 \cr
   0.9 &  -12.1 &  14.8 & 300.3 &  19.0 \cr
   1.7 &  -34.2 &  12.8 & 282.0 &  16.4 \cr
   2.9 &  -46.7 &  11.9 & 269.9 &  15.2 \cr
   4.5 &  -62.5 &  17.2 & 229.2 &  22.0 \cr
   6.1 &  -84.1 &  25.3 & 233.2 &  32.3 \cr
   8.1 &  -67.6 &  38.7 & 308.4 &  49.8 \cr
  11.5 &  -94.8 &  40.5 & 225.5 &  51.8 \cr
  17.1 & -124.6 &  43.7 & 204.3 &  56.0 \cr
  27.4 &  -99.3 &  75.0 & 200.0 &  65.6 \cr
}
\endtable

\begintable{7}
\caption{NGC~2563 PA 90 minor axis }
\halign{%
\rm#\hfil& \qquad\rm#\hfil& \quad\rm\hfil#& \quad\rm\hfil# &\quad\rm\hfil# \cr
  R($^{\prime\prime}$) &  $v$        &  $\Delta v$ & $\sigma$    &$\Delta \sigma $\cr
\noalign{\vskip 10pt}
  -11.2 &-45.2& 60.4&     173.2&    77.8\cr   
  -6.1  &-8.6&     33.9&     153.5&    43.9\cr   
  -3.0  &-44.5&     26.3&     251.0&    33.8\cr   
  -1.2  &-28.9&     18.4&     248.0&    23.6\cr   
  0.4  & -17.4&     18.1&     273.0&    23.3\cr   
   2.0 &  -67.0&      16.1&      272.3&     20.7\cr
   3.9 &-40.8&    24.0   &    233.6    &   30.8  \cr  
   6.9 &-42.2&    48.9&    222.1    &   62.8  \cr  
}
\endtable
 
 \begintable{8}
\caption{NGC~3078 PA 177 major axis }
\halign{%
\rm#\hfil& \qquad\rm#\hfil& \quad\rm\hfil#& \quad\rm\hfil# &\quad\rm\hfil# \cr
  R($^{\prime\prime}$) &  $v$        &  $\Delta v$ & $\sigma$    &$\Delta \sigma $\cr
\noalign{\vskip 10pt}
  -34.4&      55.9&      46.5&  172.86&    77.37\cr
  -27.1&      35.9&      62.9&  174.09&    87.04\cr
  -21.4&      26.3&     62.4&   281.2  &     81.2\cr
  -15.6&      72.2&    45.4&   266.5&    59.0\cr
  -10.8&      64.9&     25.8&    274.1&     33.6\cr
  -7.2&      101.9&     21.1&    254.6&    27.5\cr
  -4.9&      78.9&     15.3&    244.9&    19.8\cr
  -3.3&      71.4&     11.3&    252.9&    14.7\cr
  -1.7&      40.4&     11.1&    271.6&   14.4\cr
 -0.1&     5.0&     10.8   &    272.1&    14.0\cr
   1.5&     -38.2&     11.7&    246.7&    15.2\cr
   3.1&     -57.6&     15.7&    291.8&    20.5\cr
   4.7&    -76.1&     18.9&    225.1&    24.5\cr
   6.7&    -91.1&     20.0&    231.1&    26.0\cr
   9.8&    -70.1&     24.2&    226.38&    42.65\cr
   14.9&    -82.4&     45.8&   269.43&    85.34\cr
   23.1&    -92.2&     56.5&    257.9&    129.0\cr
   34.4&    -111.5&     36.9&    105.2&    47.7\cr
 }
\endtable

 \begintable{9}
\caption{NGC~3258 PA 75 major axis }
\halign{%
\rm#\hfil& \qquad\rm#\hfil& \quad\rm\hfil#& \quad\rm\hfil# &\quad\rm\hfil# \cr
  R($^{\prime\prime}$) &  $v$        &  $\Delta v$ & $\sigma$    &$\Delta \sigma $\cr
\noalign{\vskip 10pt}
-20.4 &  24.1 &  33.0 & 154.6 &  44.7 \cr
-13.4 &  42.6 &  25.1 & 172.9 &  34.2 \cr
 -8.6 &  66.2 &  20.0 & 210.6 &  27.5 \cr
 -6.0 &  25.8 &  24.0 & 256.0 &  32.6 \cr
 -4.4 &  46.4 &  18.5 & 302.5 &  24.6 \cr
 -2.8 &  32.9 &  12.1 & 282.7 &  16.3 \cr
 -1.2 &  54.4 &  12.0 & 316.5 &  15.9 \cr
  0.0 &  -0.9 &  13.4 & 281.2 &  18.0 \cr
  1.2 & -27.9 &  10.5 & 250.4 &  14.2 \cr
  2.8 & -41.5 &  12.9 & 241.4 &  17.6 \cr
  4.7 & -37.0 &  13.8 & 199.9 &  19.0 \cr
  7.2 & -45.6 &  20.3 & 230.6 &  27.8 \cr
 11.0 & -36.1 &  22.0 & 161.4 &  29.8 \cr
 16.8 & -48.5 &  50.6 & 140.0 &  49.7 \cr
}
\endtable
\bigskip
 \begintable{10}
\caption{NGC~3923 PA 50 major axis }
\halign{%
\rm#\hfil& \qquad\rm#\hfil& \quad\rm\hfil#& \quad\rm\hfil# &\quad\rm\hfil# \cr
  R($^{\prime\prime}$)  &  $v$        &  $\Delta v$ & $\sigma$    &$\Delta \sigma $\cr
\noalign{\vskip 10pt}
 -30.1 &  -44.2 &  43.5 & 262.6 &  96.7 \cr
 -24.1 &    9.8 &  26.4 & 246.5 &  34.1 \cr
 -17.9 &   20.9 &  20.5 & 258.7 &  30.8 \cr
 -12.8 &    8.7 &  17.4 & 248.9 &  21.5 \cr
  -9.7 &   24.1 &  19.8 & 247.6 &  25.7 \cr
  -6.3 &    4.5 &  13.2 & 252.4 &  13.3 \cr
  -3.6 &    5.4 &  10.0 & 245.6 &  14.7 \cr
  -2.0 &   -9.7 &  11.0 & 261.9 &  14.1 \cr
  -0.4 &  -13.8 &   9.7 & 252.6 &  12.4 \cr
   1.2 &   -8.5 &  10.8 & 243.4 &  13.9 \cr
   3.2 &   -8.1 &  10.8 & 245.1 &  13.8 \cr
   6.3 &    6.9 &  12.2 & 233.9 &  15.7 \cr
  11.1 &   25.7 &  15.0 & 237.5 &  19.2 \cr
  16.8 &   -4.5 &  25.3 & 282.4 &  32.4 \cr
  22.5 &   28.9 &  30.4 & 224.1 &  38.9 \cr
  28.2 &   53.8 &  38.1 & 166.7 &  49.2 \cr
}
\endtable
\bigskip

\begintable{11}
\caption{NGC~4526 PA 113 major axis }
\halign{%
\rm#\hfil& \qquad\rm#\hfil& \quad\rm\hfil#& \quad\rm\hfil# &\quad\rm\hfil# \cr
  R($^{\prime\prime}$) &  $v$        &  $\Delta v$ & $\sigma$    &$\Delta \sigma $\cr
\noalign{\vskip 10pt}
-135.4 &  216.0 &  50.8 & 124.8 &  81.0 \cr
-116.7 &  183.1 &  47.0 &  28.1 &  56.8 \cr
-101.7 &  251.6 &  58.3 &   0.9 & 116.9 \cr
 -88.7 &  266.4 &  34.5 &  32.8 &  47.6 \cr
 -76.5 &  256.2 &  25.3 &  89.2 &  32.1 \cr
 -64.3 &  202.9 &  19.2 &  71.3 &  24.4 \cr
 -53.5 &  245.3 &  14.4 &  72.0 &  18.4 \cr
 -45.5 &  200.6 &  26.4 & 163.5 &  36.0 \cr
 -39.4 &  155.4 &  33.1 & 198.8 &  45.6 \cr
 -33.7 &  132.4 &  22.8 & 162.2 &  31.1 \cr
 -28.5 &  153.0 &  30.2 & 193.9 &  41.6 \cr
 -23.4 &  132.2 &  17.2 & 179.7 &  23.6 \cr
 -15.6 &  144.5 &  11.6 & 167.6 &  15.8 \cr
 -11.9 &  175.1 &   8.9 & 156.7 &  12.0 \cr
  -7.9 &  170.2 &  13.7 & 186.0 &  18.9 \cr
  -7.1 &  175.0 &  11.1 & 178.9 &  15.3 \cr
  -6.3 &  179.4 &   9.9 & 173.5 &  13.5 \cr
  -5.5 &  165.2 &   9.3 & 182.1 &  12.7 \cr
  -4.7 &  154.9 &   8.9 & 199.6 &  12.2 \cr
  -3.9 &  127.3 &   8.2 & 192.0 &  11.3 \cr
  -3.1 &  112.8 &   7.8 & 205.1 &  10.7 \cr
  -2.3 &   86.1 &   7.2 & 216.2 &  10.0 \cr
  -1.4 &   51.4 &   7.8 & 253.6 &  10.7 \cr
  -0.6 &   28.5 &   7.8 & 245.8 &  10.8 \cr
   0.2 &    0.0 &   7.5 & 222.0 &  10.4 \cr
   1.0 &  -23.6 &   7.4 & 243.8 &  10.2 \cr
   1.8 &  -47.1 &   8.8 & 219.9 &  12.1 \cr
   2.6 &  -70.9 &   8.4 & 231.2 &  11.6 \cr
   3.4 &  -69.8 &   9.1 & 219.0 &  12.5 \cr
   4.2 & -118.7 &   7.8 & 186.6 &  10.8 \cr
   5.0 & -136.1 &   9.4 & 216.3 &  13.0 \cr
}
\endtable

\begintable{12}
\caption{NGC~4526 PA 113 major axis -- {\it continued}}
\halign{%
\rm#\hfil& \qquad\rm#\hfil& \quad\rm\hfil#& \quad\rm\hfil# &\quad\rm\hfil# \cr
 R($^{\prime\prime}$) &  $v$        &  $\Delta v$ & $\sigma$    &$\Delta \sigma $\cr
\noalign{\vskip 10pt}
   5.8 & -143.4 &   9.3 & 188.8 &  12.8 \cr
   6.7 & -168.2 &  10.7 & 220.9 &  14.7 \cr
   7.5 & -147.6 &  11.4 & 190.4 &  15.7 \cr
   8.3 & -155.1 &  12.0 & 207.7 &  16.5 \cr
   9.1 & -186.2 &  10.5 & 142.5 &  14.2 \cr
   9.9 & -147.8 &  16.3 & 197.3 &  22.4 \cr
  10.7 & -144.4 &  16.6 & 189.1 &  22.9 \cr
  12.3 & -114.4 &  17.6 & 192.6 &  24.3 \cr
  13.1 & -141.6 &  20.4 & 168.0 &  27.9 \cr
  16.8 & -140.1 &  16.5 & 174.8 &  22.6 \cr
  18.4 & -161.8 &  18.0 & 180.4 &  24.8 \cr
  20.4 & -106.8 &  15.9 & 155.0 &  21.7 \cr
  22.8 &  -83.8 &  18.3 & 163.0 &  25.0 \cr
  26.0 & -101.4 &  16.5 & 183.3 &  22.6 \cr
  30.1 & -158.2 &  21.7 & 192.8 &  29.9 \cr
  34.1 & -170.5 &  16.3 & 123.6 &  21.9 \cr
  38.2 & -195.7 &  23.3 & 115.2 &  30.8 \cr
  42.2 & -189.6 &  22.5 & 138.6 &  30.2 \cr
  46.3 & -230.0 &  29.5 & 135.1 &  39.8 \cr
  51.5 & -250.4 &  21.1 & 120.8 &  28.0 \cr
  57.6 & -192.8 &  23.3 & 125.9 &  31.1 \cr
  62.9 & -292.2 &  42.3 & 122.9 &  56.2 \cr
  68.5 & -238.9 &  22.5 &  74.8 &  28.7 \cr
  75.1 & -280.4 &  27.4 &   0.4 &  36.9 \cr
  81.1 & -270.1 &  27.8 & 187.0 &  54.8 \cr
  87.2 & -273.3 &  21.4 &  53.0 &  29.4 \cr
  94.0 & -247.2 &  27.2 &  13.1 &  83.2 \cr
 101.3 & -251.5 &  30.2 &  60.0 &  27.6 \cr
 109.0 & -303.7 &  34.4 &  60.8 &  20.6 \cr
 118.9 & -252.1 &  29.6 &  60.8 &  39.1 \cr
 131.1 & -288.5 &  57.2 &  60.0 &  62.7 \cr
}
\endtable

\begintable{13}
\caption{NGC~4753 PA 80 major axis }
\halign{%
\rm#\hfil& \qquad\rm#\hfil& \quad\rm\hfil#& \quad\rm\hfil# &\quad\rm\hfil# \cr
 R($^{\prime\prime}$) &  $v$        &  $\Delta v$ & $\sigma$    &$\Delta \sigma $\cr
\noalign{\vskip 10pt}
 -70.0 & -186.6 &  95.8 &  80.1 &  99.2 \cr
 -58.4 & -163.9 &  58.0 &  99.1 &  74.8 \cr
 -48.7 & -129.9 &  33.9 & 100.1 &  34.6 \cr
 -41.0 & -129.0 &  23.0 &  93.7 &  29.9 \cr
 -32.9 & -152.7 &  26.5 & 124.0 &  35.4 \cr
 -26.1 & -144.6 &  20.1 & 100.0 &  20.6 \cr
 -21.3 & -134.1 &  26.5 & 142.3 &  35.7 \cr
 -16.8 & -131.3 &  19.6 & 154.2 &  26.6 \cr
 -11.9 & -118.7 &  21.4 & 171.1 &  29.3 \cr
  -7.3 &  -98.4 &  13.5 & 169.3 &  18.4 \cr
  -5.8 & -104.3 &  12.0 & 154.8 &  16.4 \cr
  -3.9 & -101.9 &   9.3 & 160.6 &  12.6 \cr
  -2.7 &  -54.3 &  10.6 & 163.0 &  12.1 \cr
  -1.9 &  -31.8 &   7.1 & 169.6 &  10.9 \cr
  -1.1 &  -37.7 &   7.5 & 171.0 &  10.2 \cr
  -0.3 &  -17.0 &   6.6 & 163.8 &  10.5 \cr
   0.5 &    2.6 &   6.2 & 165.3 &   8.4 \cr
   1.3 &   21.6 &   6.7 & 172.6 &  10.3 \cr
   2.1 &   60.0 &   7.2 & 172.4 &   9.8 \cr
   3.0 &   75.5 &   9.4 & 168.4 &  12.9 \cr
   3.8 &   92.2 &  12.4 & 176.0 &  13.2 \cr
   5.0 &  105.9 &   8.6 & 173.0 &  11.0 \cr
   6.6 &  125.9 &  12.8 & 159.6 &  17.4 \cr
   9.4 &  128.0 &  14.2 & 159.4 &  19.4 \cr
  11.8 &  147.2 &  21.5 & 160.1 &  29.2 \cr
  14.2 &  155.8 &  20.4 & 131.7 &  27.4 \cr
  17.1 &  122.7 &  18.6 & 104.9 &  24.4 \cr
  21.1 &  123.8 &  17.3 & 117.8 &  23.0 \cr
  26.0 &  111.7 &  18.9 &  95.4 &  24.6 \cr
  30.8 &  125.1 &  22.8 &  93.3 &  24.4 \cr
  35.6 &  117.8 &  17.2 &  82.9 &  21.6 \cr
  41.8 &  134.6 &  34.6 & 118.9 &  33.1 \cr
  49.7 &  115.5 &  44.6 &  70.7 &  35.4 \cr
  60.5 &   90.6 &  48.7 & 119.1 &  65.8 \cr
  74.1 &  112.2 &  41.3 &  99.0 &  49.5 \cr
}
\endtable
\bigskip

\begintable{14}
\caption{NGC~4756 PA 50 major axis }
\halign{%
\rm#\hfil& \qquad\rm#\hfil& \quad\rm\hfil#& \quad\rm\hfil# &\quad\rm\hfil# \cr
 R($^{\prime\prime}$) &  $v$        &  $\Delta v$ & $\sigma$    &$\Delta \sigma $\cr
\noalign{\vskip 10pt}
 -10.5 &   47.9 &  32.3 & 210.8 &  45.3 \cr
  -4.6 &   45.2 &  22.9 & 193.3 &  32.0 \cr
  -2.4 &   11.4 &  15.3 & 207.8 &  21.4 \cr
  -0.8 &   20.1 &  12.9 & 206.6 &  18.1 \cr
   0.8 &  -18.2 &  11.4 & 201.3 &  15.9 \cr
   2.4 &  -29.3 &  14.8 & 194.2 &  20.6 \cr
   7.4 &  -22.9 &  17.5 & 212.5 &  24.9 \cr
}
\endtable
\bigskip

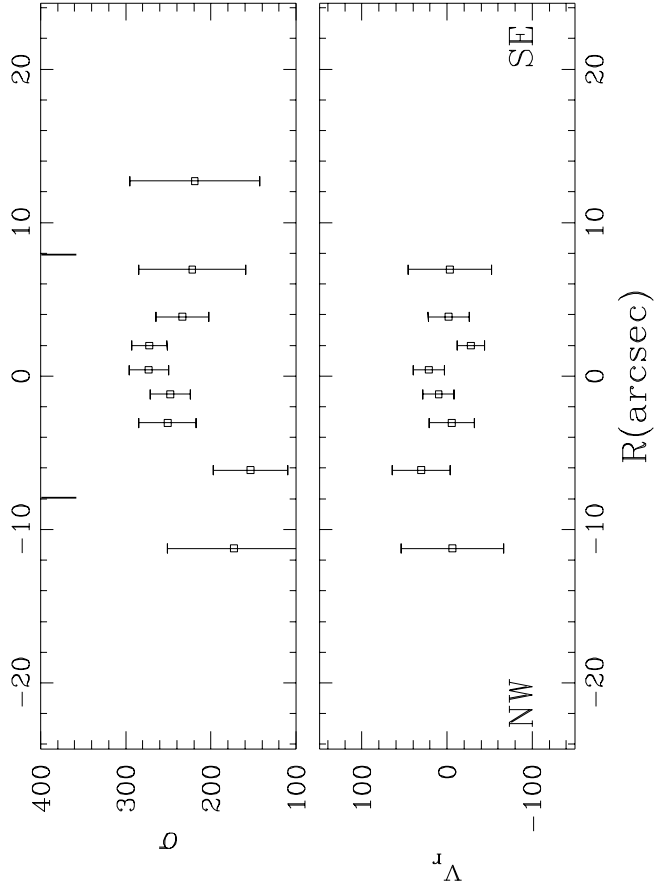
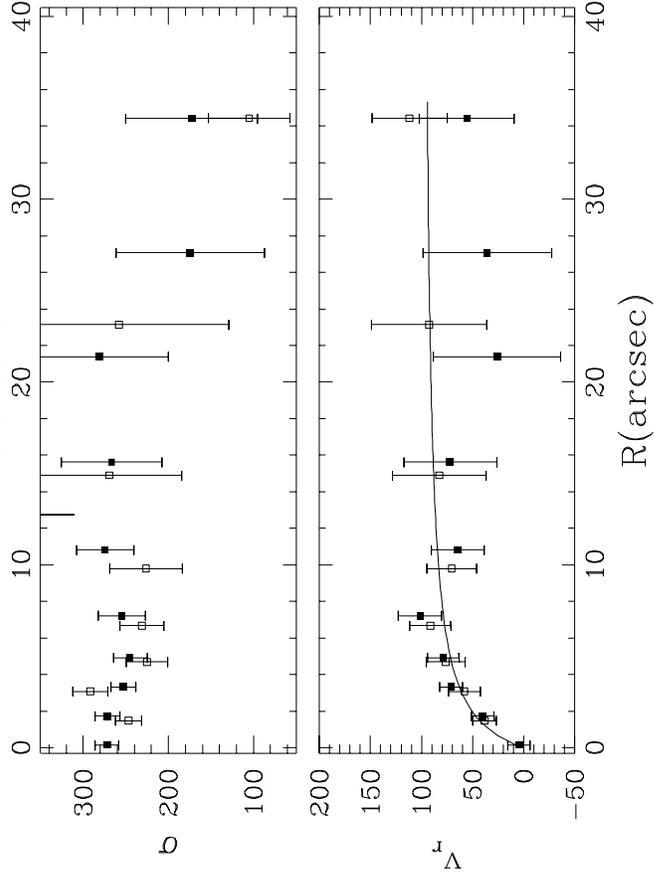
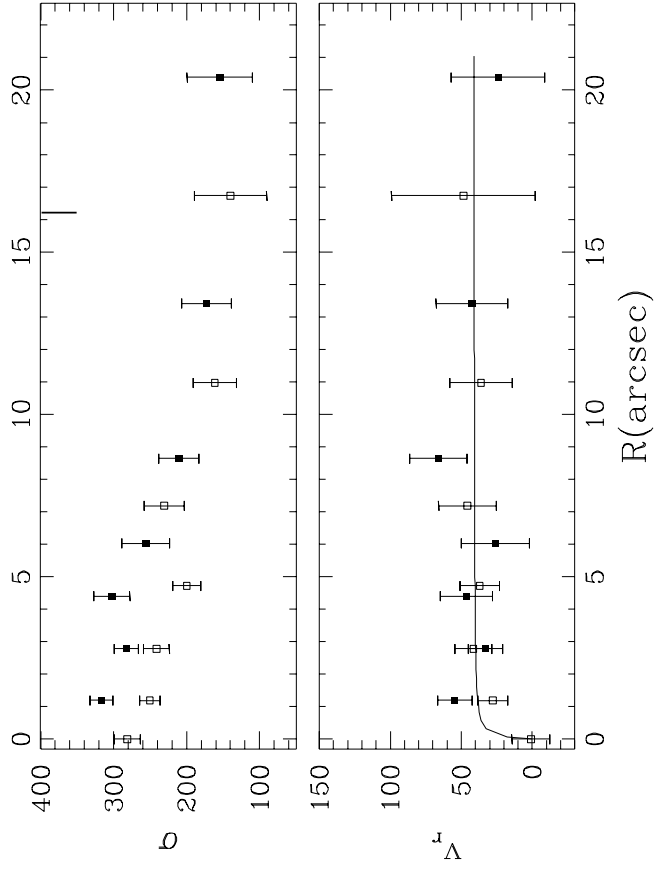
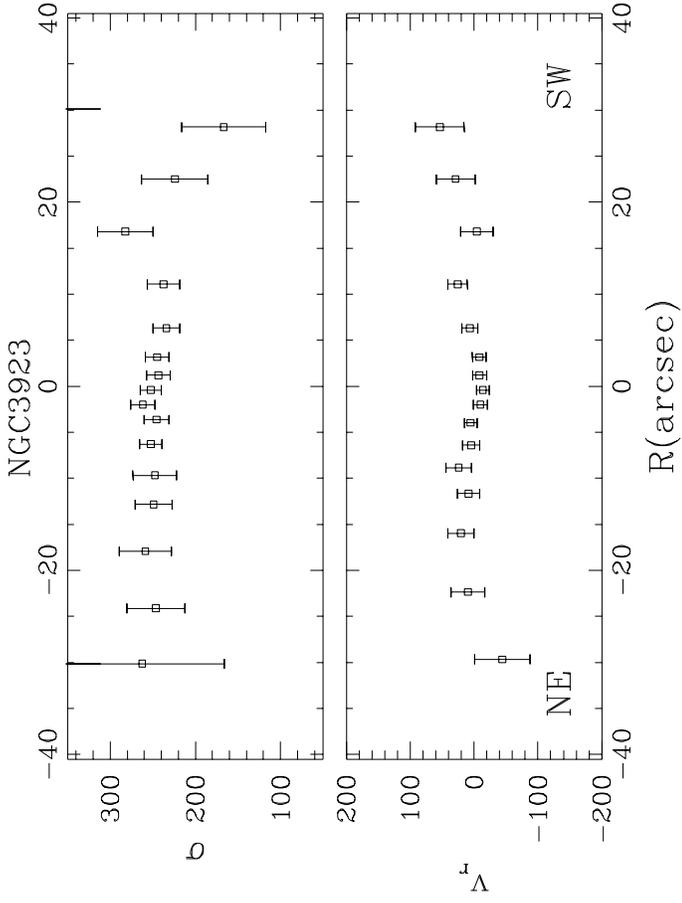
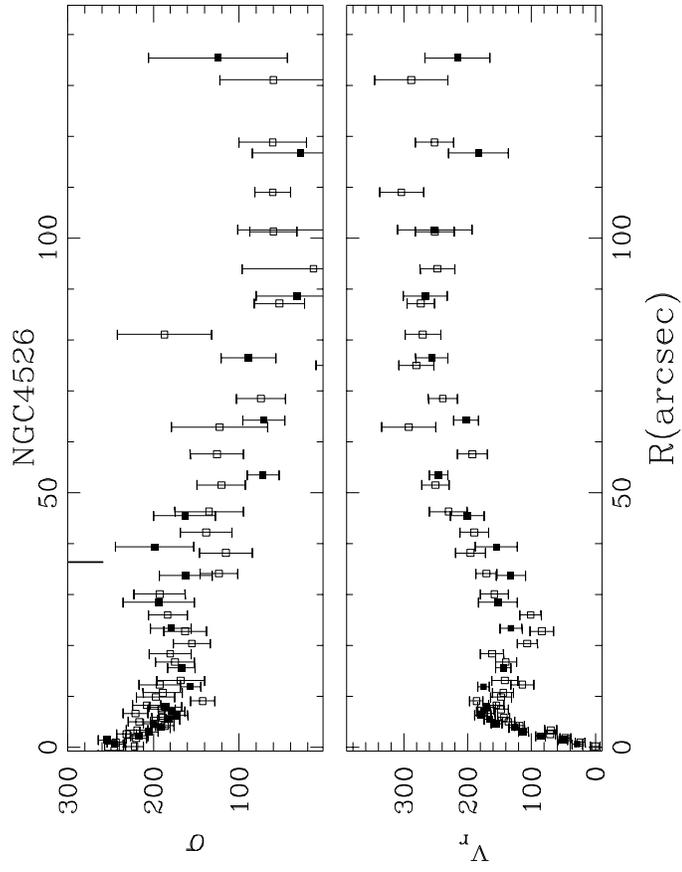
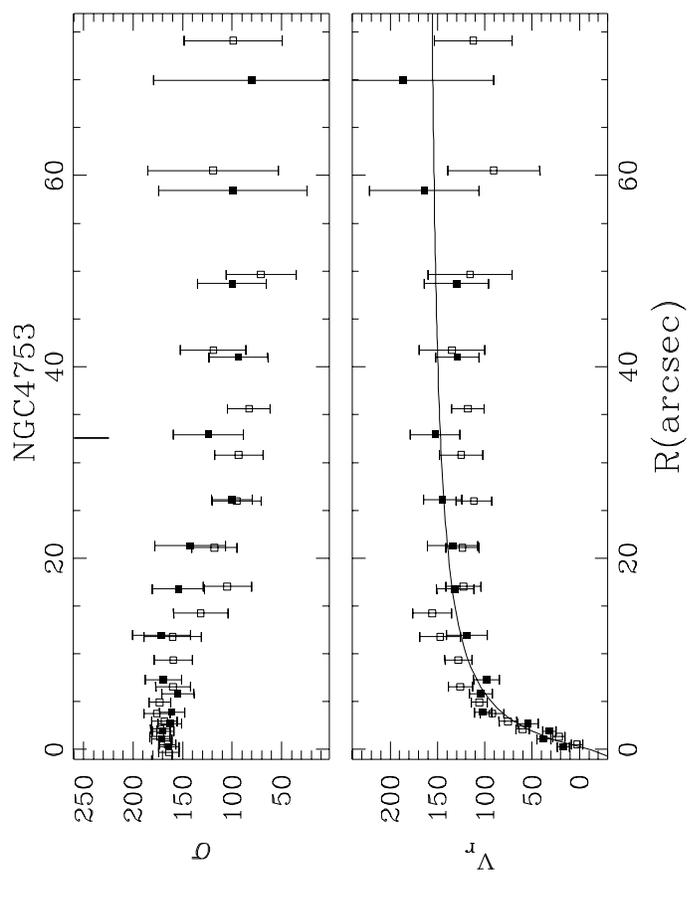
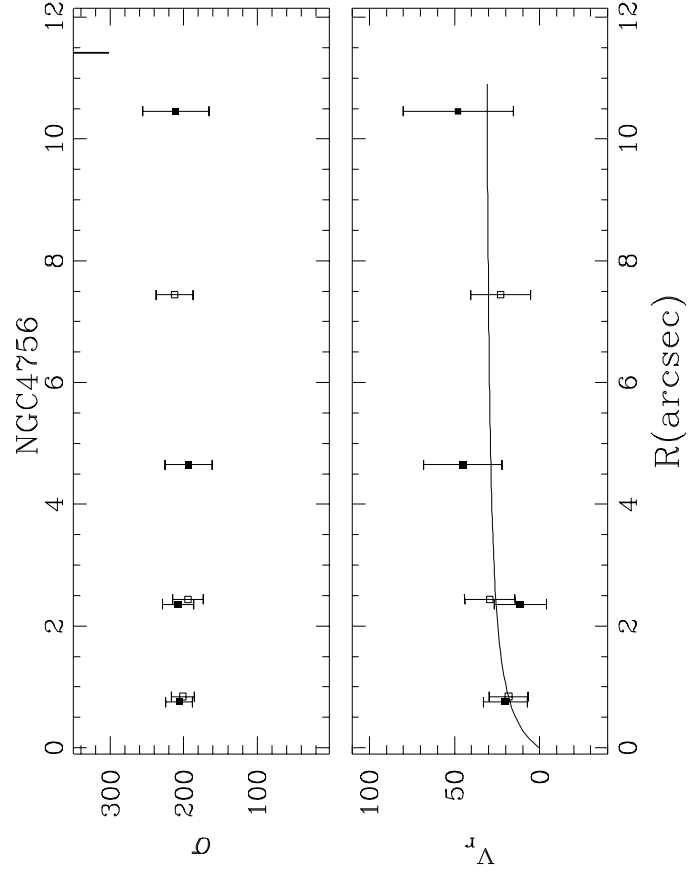
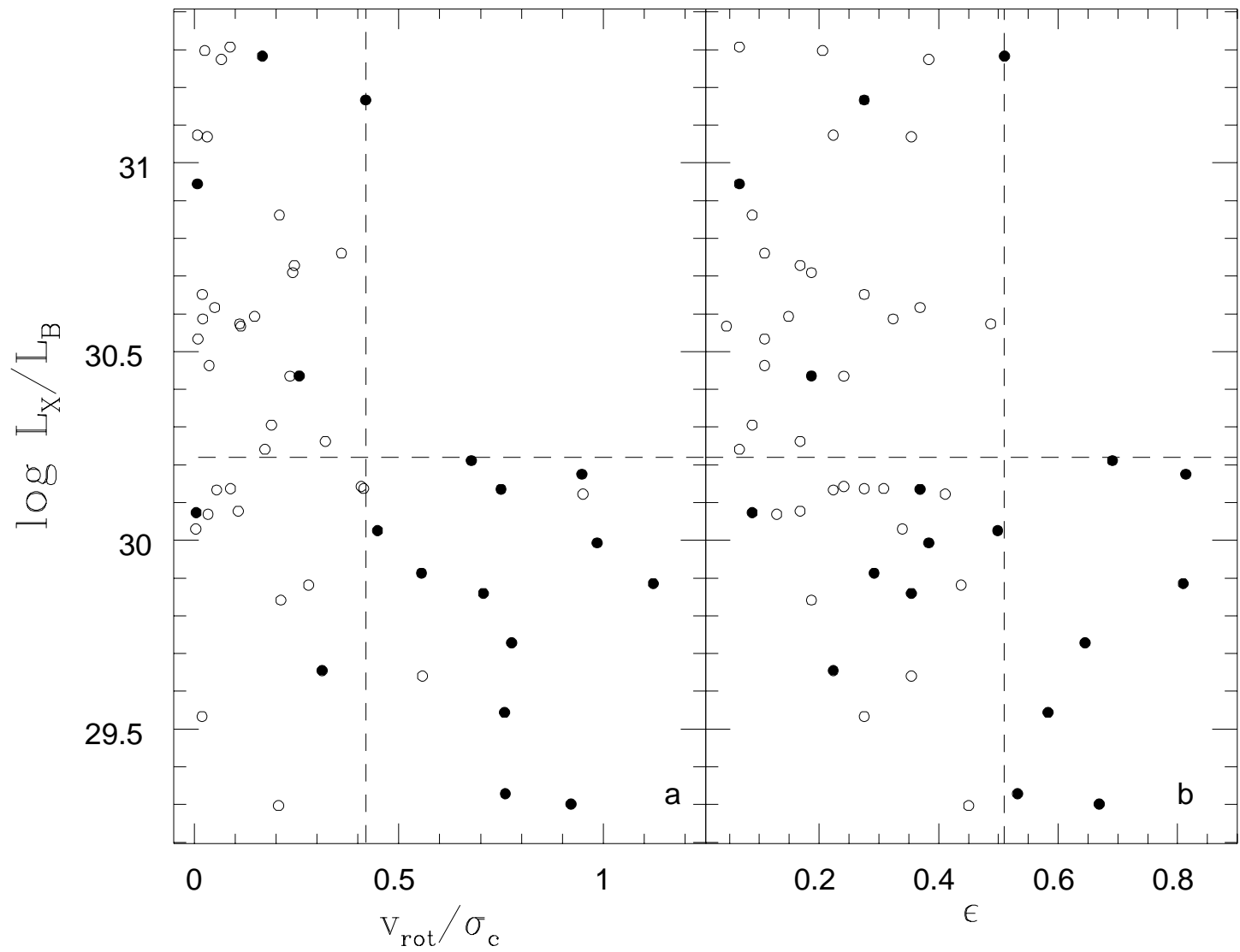
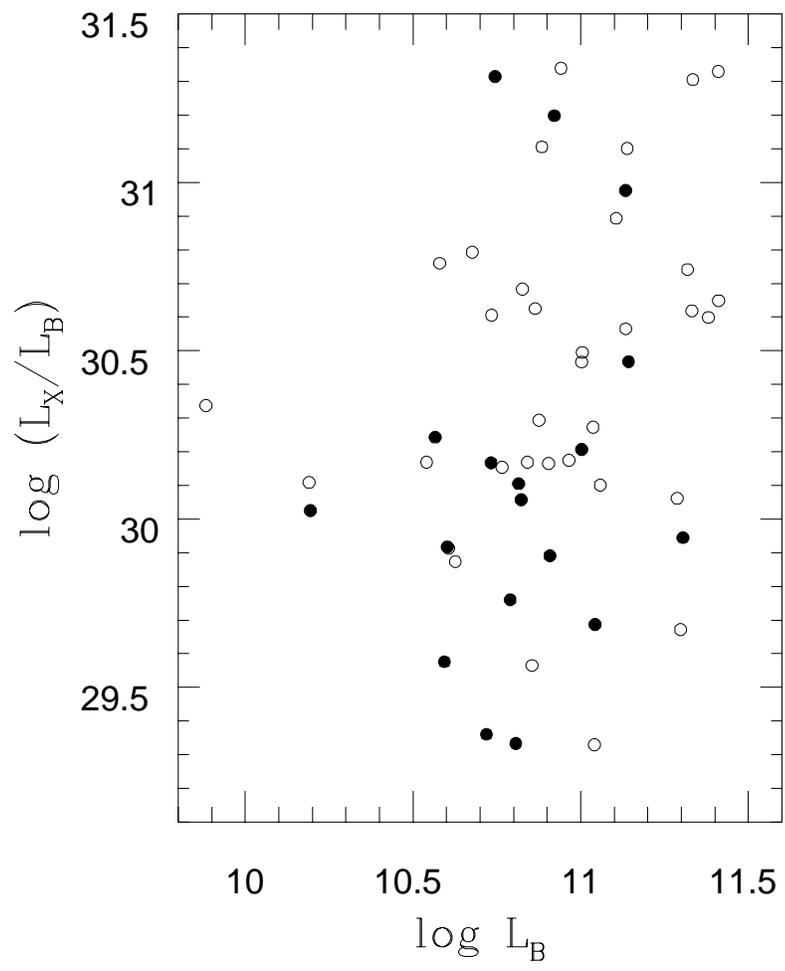
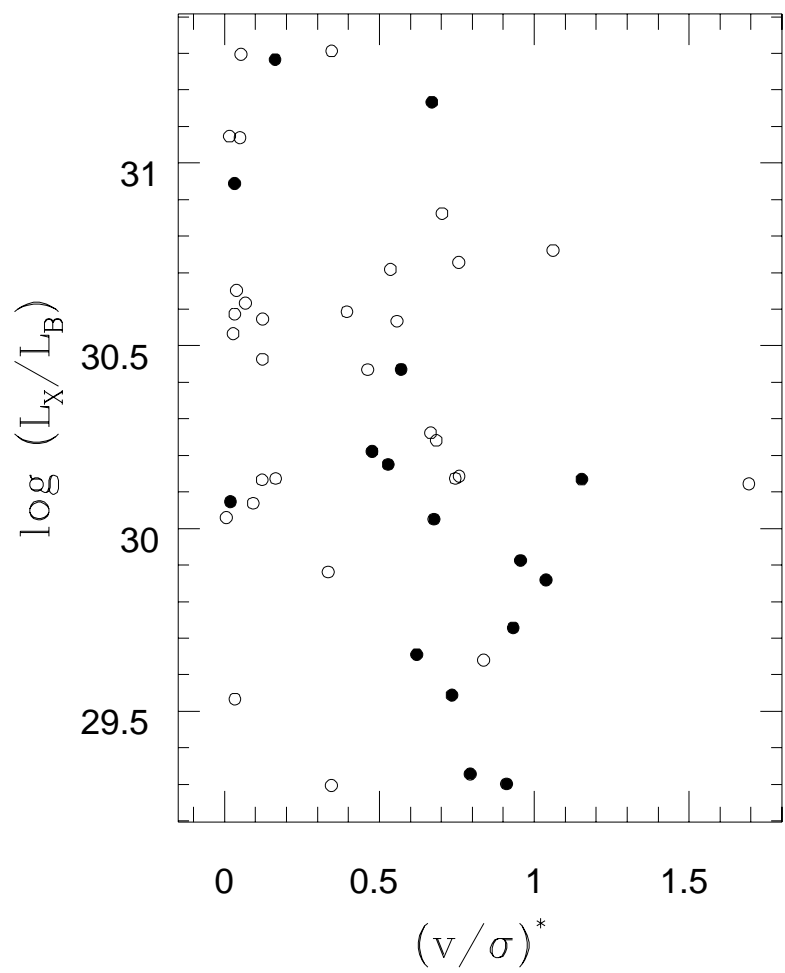
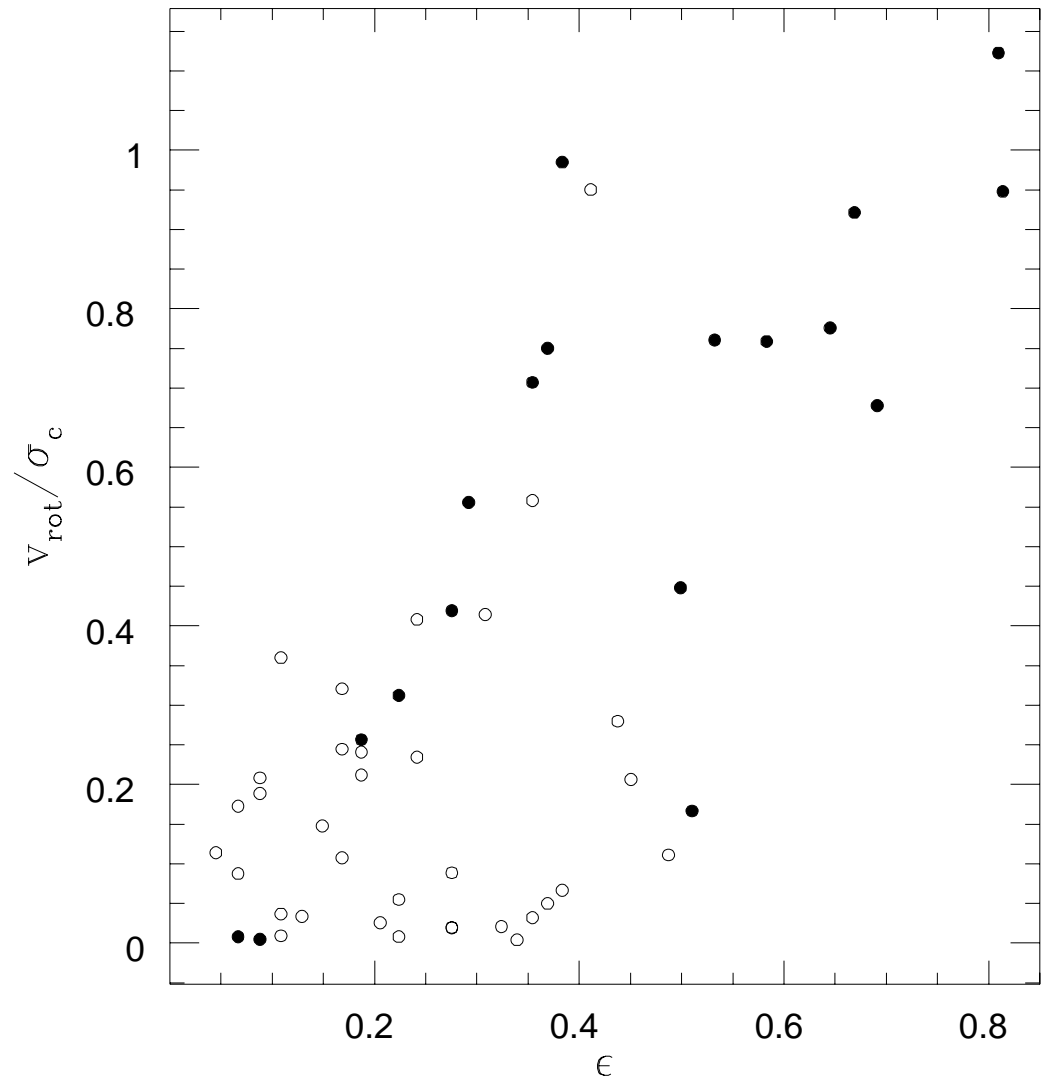
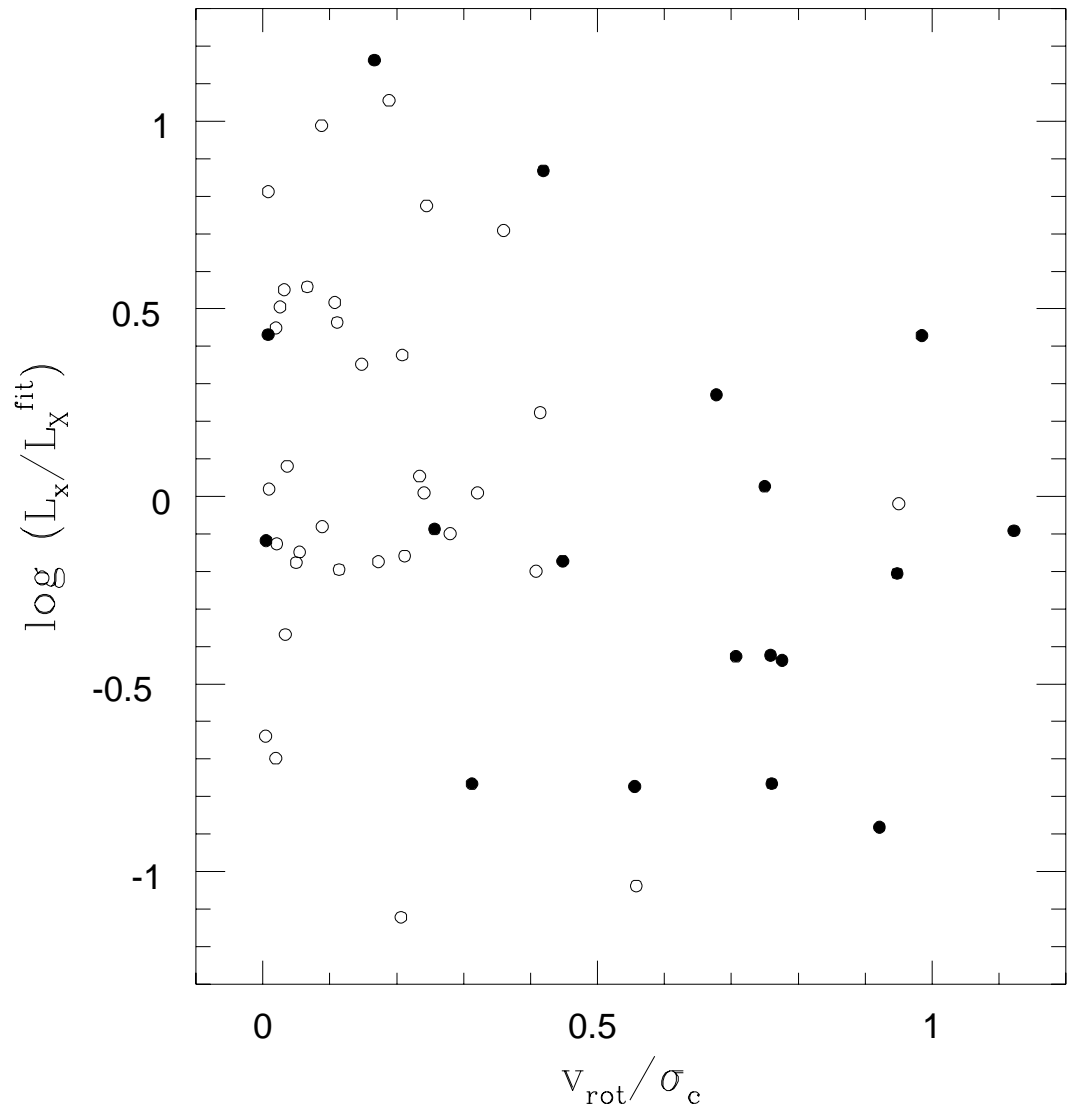
\bye